\documentclass[journal]{IEEEtran}

\pdfoutput=1

\usepackage{color}
\usepackage{amsmath}
\usepackage{graphicx}
\usepackage{booktabs}
\usepackage{tabularx}
\usepackage{amssymb} 
\usepackage{gensymb}
\usepackage{multirow}
\usepackage[table]{xcolor}
\usepackage{subfig}
\usepackage[table]{xcolor}
\usepackage{array}
\usepackage{bm}

\usepackage{pgfplots}
\pgfplotsset{compat=newest}

\newcolumntype{C}[1]{>{\centering\arraybackslash}m{#1}}
\newcolumntype{R}[1]{>{\raggedright\arraybackslash}m{#1}}
\newcolumntype{L}[1]{>{\raggedleft\arraybackslash}m{#1}}

\usepackage[bookmarks=false,colorlinks=true,citecolor=blue]{hyperref}

\def\fmd{{f_\text{mD}}}
\def\fdmax{{f^\mathrm{D}_\text{max}}}

\begin{document}
%
\title{Toward Unobtrusive In-home Gait Analysis \\ Based on Radar Micro-Doppler Signatures}
%
%
\author{Ann-Kathrin~Seifert,~\IEEEmembership{Student Member,~IEEE,}
        Moeness~G.~Amin,~\IEEEmembership{Fellow,~IEEE,}
        and~Abdelhak~M.~Zoubir,~\IEEEmembership{Fellow,~IEEE}%
\thanks{Copyright (c) 2019 IEEE. Personal use of this material is permitted. However, permission to use this material for any other purposes must be obtained from the IEEE by sending an email to pubs-permissions@ieee.org.

The work by M.~G.~Amin is supported by the Alexander von Humboldt Foundation, Bonn, Germany.
	
A.-K.~Seifert and A.~M.~Zoubir are with the Signal Processing Group at Technische Universit\"at Darmstadt, Darmstadt, Germany (correspondence e-mail: seifert@spg.tu-darmstadt.de).

M.~G.~Amin is with the Center for Advanced Communications at Villanova University, Villanova, PA, USA. 
}}
\maketitle

\begin{abstract}
\textit{Objective:} In this paper, we demonstrate the applicability of radar for gait classification with application to home security, medical diagnosis, rehabilitation and assisted living. Aiming at identifying changes in gait patterns based on radar micro-Doppler signatures, this work is concerned with solving the intra motion category classification problem of gait recognition. 
\textit{Methods:} New gait classification approaches utilizing physical features, subspace features and sum-of-harmonics modeling are presented and their performances are evaluated using experimental K-band radar data of four test subjects. Five different gait classes are considered for each person, including normal, pathological and assisted walks. 
\textit{Results:} The proposed approaches are shown to outperform existing methods for radar-based gait recognition which utilize physical features from the cadence-velocity data representation domain as in this paper. The analyzed gait classes are correctly identified with an average accuracy of 93.8\%, where a classification rate of 98.5\% is achieved for a single gait class. When applied to new data of another individual a classification accuracy on the order of 80\% can be expected.
\textit{Conclusion:} Radar micro-Doppler signatures and their Fourier transforms are well suited to capture changes in gait. Five different walking styles are recognized with high accuracy. 
\textit{Significance:} Radar-based sensing of gait is an emerging technology with multi-faceted applications in security and health care industries. We show that radar, as a contact-less sensing technology, can supplement existing gait diagnostic tools with respect to long-term monitoring and reproducibility of the examinations.

\end{abstract}


%
\IEEEpeerreviewmaketitle


\section{Introduction}

\IEEEPARstart{R}{ecently}, radar has received much attention in civilian applications, most notably, in automotive and health care industries. Specifically, the applications of radar technology in home security, elderly care, and medical diagnosis have emerged to be front and center in indoor human monitoring \cite{Che14,Ami16,Ami17}. These include fall motion detection, classifications of daily activities, and vital sign monitoring. The considerable rise in radar indoor applications and smart homes is credited to its safety, reliability, and ability to serve as an effective device for contact-less motion monitoring of subjects in the surrounding settings and environments, while preserving privacy. Other non-wearable sensing modalities for indoor human monitoring include infrared reflective light \cite{Hag10}, refractive light, video cameras, and in-ground force platforms (for an overview see e.g.~\cite{Mur14}). However, visual perception or video recordings of human motions can easily be disturbed by occlusions, lighting conditions and clothing.

Low-cost Doppler radars have widely been used for detection \cite{Wan14,Su15,Kim15,Cle15,Jok18}, identification \cite{Van18}, classification \cite{Kim09,Bjoe15,Jok17}, and recognition of human motions \cite{Ric15}. However, most of these works are concerned with the discrimination between different classes of motions, and as such, they consider inter motion category classification problems. A nominal example is discriminating between running, walking, crawling, creeping, sitting, bending and falling. Yet, little thought has been given to study the intra motion category classification problem, i.e., discerning variations within one motion class.

In this paper, we focus on classifying gait within its class. Gait analysis plays a key role in medical diagnosis, biomedical engineering, sports medicine, physiotherapy and rehabilitation \cite{Mur14}. Constant monitoring of changes in gait aids in assessing recovery from body injury. Further, it enables early diagnosis of different diseases, including multiple sclerosis, Parkinson's and cardiopathies, and facilitates studying the course of disease for designing adequate treatment \cite{Mur14}. For these reasons, it is important to detect gait abnormalities and monitor alterations in walking patterns over time. However, detailed gait analysis and proper assessments of walking aids can prove difficult for physicians, health care providers and nursing staff. Thorough clinical gait studies are often time-consuming, costly and lack reproducibility \cite{Sim04}. That is why we seek a contact-less sensing technology to empower, and not necessarily replace, naked eye gait examination, with the goal of achieving an expedited, more accurate and more efficient gait diagnosis. 

Gait abnormalities also include using assisted walking devices, which are used by a great number of seniors, and include canes and walkers. These devices can compensate for decrements in balance, gain mobility and overcome the fear of falling. It is noted that in 2011, 8.5 million U.S.~seniors aged 65 and older reported having assistive walking devices, with a cane being most commonly used by two thirds of the elderly \cite{Gel15}. In this regard, the correct use of mobility devices becomes essential to guarantee optimal support and avoid postural deformities or physical impairments with the purpose of re-establishing a normal gait. 

Using electromagnetic sensing modality, we consider classifying different walking styles, and thus demonstrate the effectiveness of radar in detecting subclasses of gait abnormalities. We show that radar can present a viable, convenient and contact-less supplement or alternative to other sensors, in particular, wearable devices (for an overview see e.g.~\cite{Tao12}), which have widely been used to study gait (for recent works see e.g.~\cite{Pha17,Ren17,Bro15}). As opposed to prior radar-based gait classification methods, which consider walks with and without arm swinging \cite{Mob09,Tiv10,Tiv15}, or different speeds of walking \cite{Cle15,Ric15}, we focus on detecting differences in the lower limbs kinematics.

For this purpose, we devise a new approach based on predefined features for classifying gaits, where normal, pathological and assisted walks are considered. We analyze two types of limping gait, where one or both legs are not swinging normally. Further, we consider two different synchronization styles between the cane and the legs and their effects on the detection of walking aids; a subject that has gained increased interest in the latest past \cite{Gur17,Sey18}. 

In addition to physical-based feature extractions, the paper considers automatic data-driven learning via subspace analysis. Applied in the cadence domain, we show that features based on principal component analysis (PCA) lead to desirable results that outperform those based on kinematic modeling. Considering five different gait classes, a normal walking is correctly detected in 94\% of the cases. Although viewed in the same category as neural networks in unsupervised feature learning, PCA does not demand the same level of computations as deep learning approaches \cite{Jok16,Kim16,Sey18,Van18}, neither does it necessitate a very large number of (training) samples.  

In order to validate the performance of the proposed method in detecting gait asymmetries, we collected radar data of four individuals with different diagnosed gait disorders. The corresponding radar data representations reveal characteristic features that indicate gait disorders. It is shown that by applying the proposed classification method, the gait asymmetry is correctly detected with high sensitivity for three of the four test subjects.

The remainder of the paper is organized as follows. For analyzing backscattered radar data from human motions, Section~\ref{sec:signalrepresentations} briefly motivates and outlines different radar data representations, which can be utilized for feature extraction. In Section~\ref{sec:featextraction}, we propose feature extraction techniques for gait classification. Corresponding results are presented and discussed in Section~\ref{sec:results}. Conclusions are given in Section~\ref{sec:conclusion}. 
\section{Representations of Human Radar Signatures}\label{sec:signalrepresentations}

The human walk is periodic by nature, i.e., after taking two steps the course of motions is repeated, which constitutes a gait cycle \cite{Che11}. One would expect the time-domain radar return signal from a walking person to be periodic with each gait cycle. However, the periodicity information cannot be directly accessed in time-domain, because it is 'hidden' in the observed frequency of the signal. The radar return signal contains multiple time-varying Doppler shifted versions of the transmitted signal. Since these Doppler components are periodic with each gait cycle, we observe a periodically frequency-modulated signal.

A classical tool to reveal periodicities in a signal is the Fourier transform (FT). However, the FT does not depict the local frequency behavior, and as such, is not the proper analysis tool for studying the instantaneous frequency and time-dependent Doppler and micro-Doppler signal components. The individual components and their power distribution over time and frequency become visible when utilizing joint time-frequency representations (TFRs). For a walking person, the TFR of the radar backscattering depicts the main Doppler shift due to the torso's motion along with the micro-Doppler components due to swinging arms and legs. The spectrogram, which is the energetic representation of the signal's short-time Fourier transform (STFT), is the most common TFR used for analyzing radar micro-Doppler signatures.

Since the spectrogram is a windowed FT, a periodic signal will remain periodic in its TFR, with each frequency component exhibiting the same periodicity. Therefore, the periodic structure of the cyclic motion articulations of human gait persist in the time-frequency domain, with a sparser and higher power concentration compared to the time-domain description of the signals. With this property, by taking the FT along the time variable for each frequency bin in the spectrogram, we can assess how often certain Doppler shifts appear over time. The result is known as the cadence-velocity diagram (CVD) \cite{Cle15,Bjoe15,Ric15,Ote05}, where velocity is proportional to the observed Doppler shifts. For a normal gait, the fundamental frequency in the CVD represents the stride rate or cadence.

\subsection{Radar Signal Model}
Considering a mono-static radar system that transmits a sinusoidal signal \cite{Che11}
\begin{equation} 
s_t(t) = \cos{\left(2 \pi f_c t\right)}
\end{equation}
with carrier frequency $f_c$, the received radar return signal is a superposition of multiple radar scatter components, i.e.,
\begin{equation}
s_r(t) = \sum_i \rho_i \cos{\left(2 \pi \left(f_c + f^\mathrm{D}_i \right) t\right)}.
\label{Eq:recSig}
\end{equation}
Here, $\rho_i$ denotes the path loss of the $i$th scatter component and $f^\mathrm{D}_i $ is the corresponding observed Doppler shift given by 
\begin{equation}
f^\mathrm{D}_i (t) \approx - f_c \frac{2 v_i(t)}{c} \cos{\theta_i}, \quad \text{for~} v_i \ll c~\forall~i,
\label{eq:Doppler}
\end{equation}
where $v_i$ is the velocity of the $i$th target component, $\theta_i$ is the corresponding angle of motion relative to the radar line of sight (LOS) and $c$ is the propagation speed of the electromagnetic (EM) wave. Note that while the radar receiver remains static, the Doppler shifts are generally time-varying as the target changes its velocity over time. After a quadrature detector at the receiver we obtain
\begin{equation}
s(t) = \sum_i \frac{\rho_i}{2} \exp{\left(-j 2 \pi f^\mathrm{D}_i  t\right)},
\end{equation}
which is the baseband representation of the multi-component radar return signal in (\ref{Eq:recSig}) with each component having a distinct Doppler frequency. For further processing, $s(t)$ is sampled at an interval of $\Delta t = 1/f_s$, such that $s(n) = s(t)|_{t=n\Delta t}$ for $n = 0, \dots ,N-1$, where $f_s$ is the sampling frequency and $N \in \mathbb{N}$ is the total number of time samples in the discretized radar signal. For further processing, the mean of the radar return signal is removed such that
\begin{equation}\label{eg:remmean}
\tilde{s}(n) = s(n) - \frac{1}{N} \sum_{n=0}^{N-1} s(n).
\end{equation}

\subsection{Time-Frequency Representations}
\begin{figure}[!t]
	\vspace{-0.8em}
	\centering{
		\subfloat[Normal walk]{\includegraphics[clip, trim= 0 0 20 18, width=0.5\columnwidth]{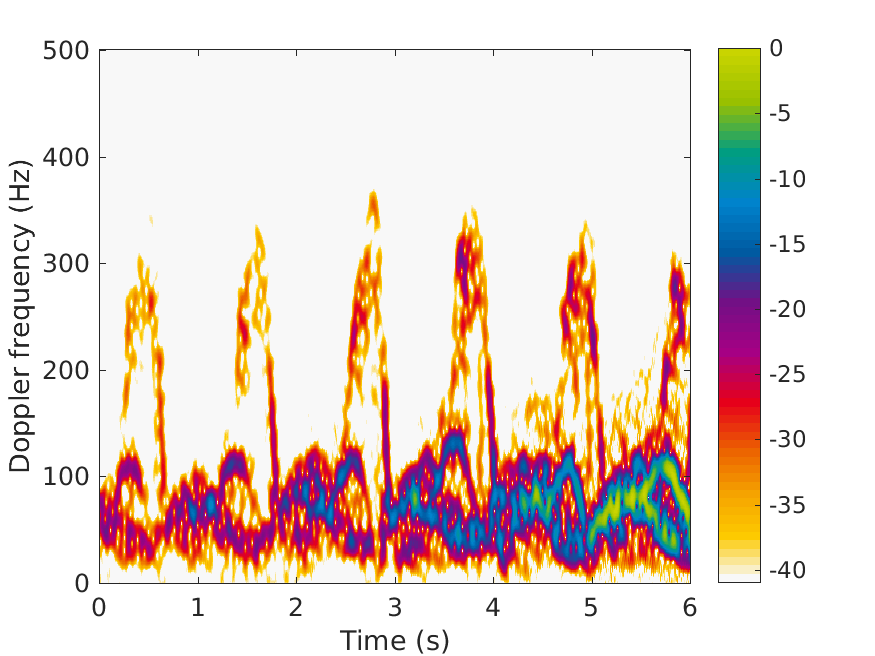}%
			\label{NW}}
		\subfloat[Normal walk]{\includegraphics[clip, trim= 0 0 20 18,width=0.5\columnwidth]{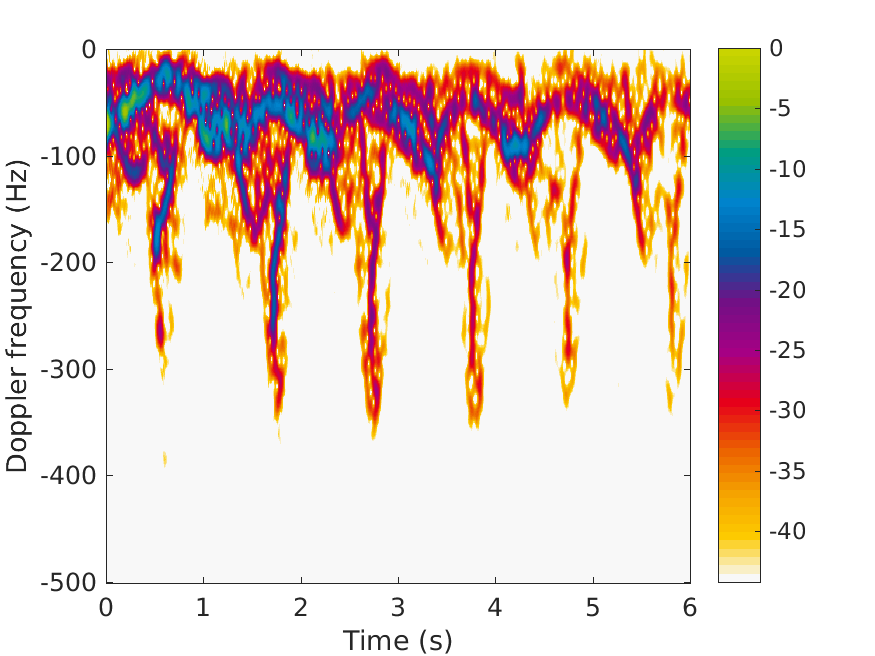}%
			\label{NWa}}\vspace{-0.7em}}

	\centering{
		\subfloat[Limping with one leg]{\includegraphics[clip, trim= 0 0 20 18,width=0.5\columnwidth]{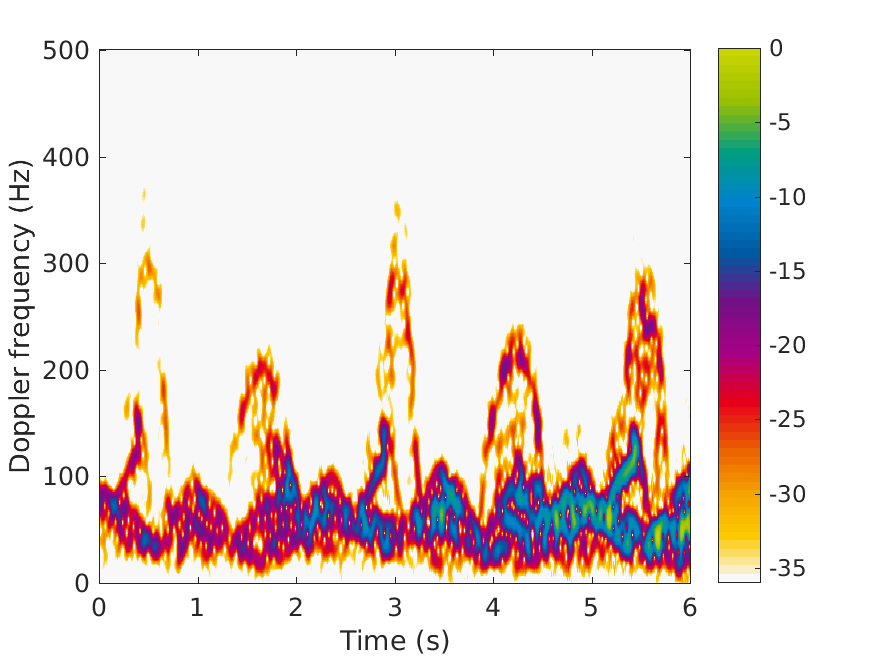}%
			\label{L1}}
		\subfloat[Limping with one leg]{\includegraphics[clip, trim= 0 0 20 18,width=0.5\columnwidth]{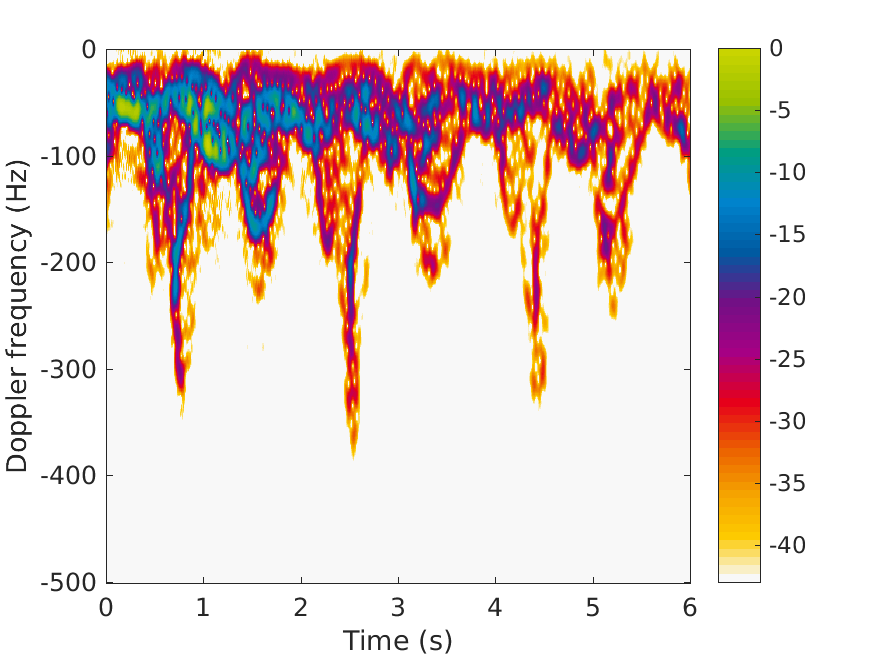}%
			\label{L1a}}\vspace{-0.7em}}

	\centering{
		\subfloat[Walking with a cane]{\includegraphics[clip, trim= 0 0 20 18,width=0.5\columnwidth]{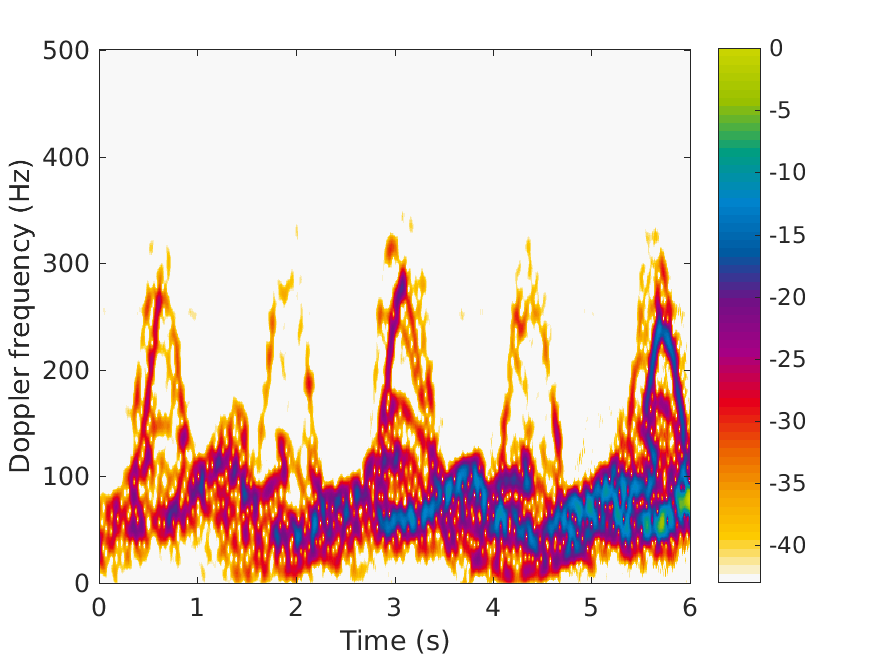}%
			\label{CW}}
		\subfloat[Walking with a cane]{\includegraphics[clip, trim= 0 0 20 18,width=0.5\columnwidth]{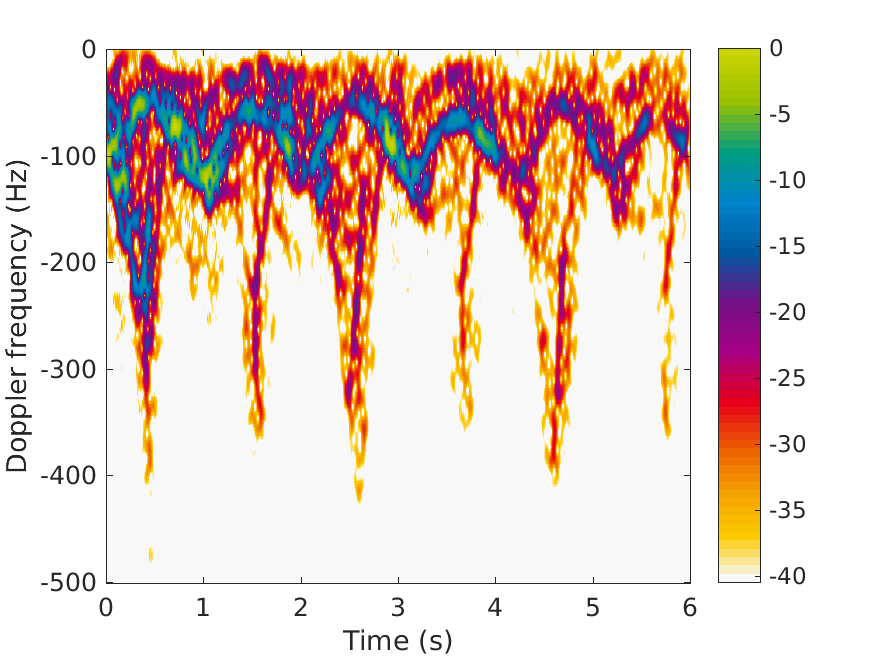}%
			\label{CWa}}\vspace{-0.7em}}

	\centering{
		\subfloat[Walking with a cane out of sync]{\includegraphics[clip, trim= 0 0 20 18,width=0.5\columnwidth]{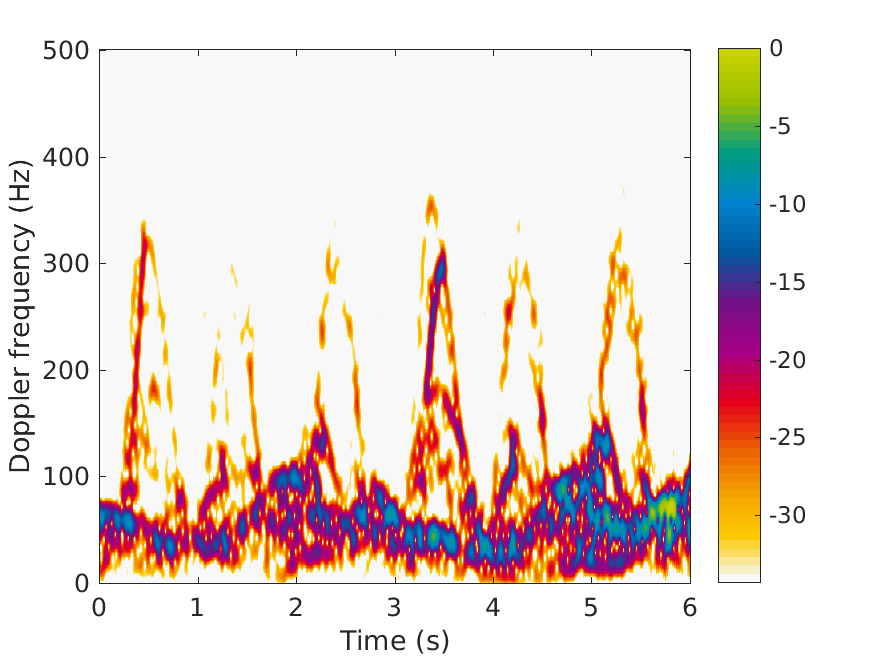}%
			\label{CWoos}}
		\subfloat[Walking with a cane out of sync]{\includegraphics[clip, trim= 0 0 20 18,width=0.5\columnwidth]{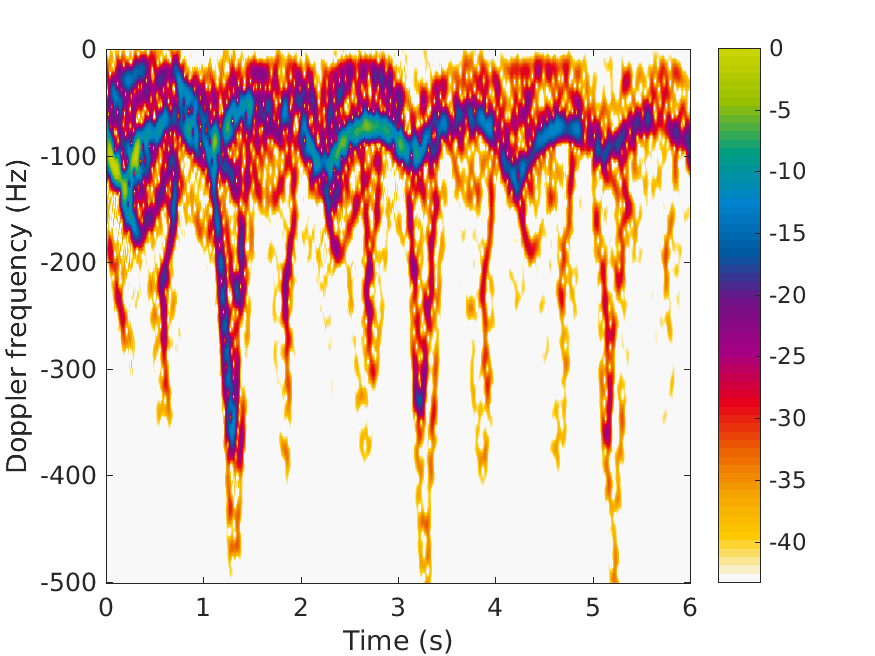}%
			\label{CWoosa}}}

	\caption{Examples of spectrograms for walking toward (left) and away from (right) the radar system. The color indicates the energy level in dB.}
	\label{fig:specs}\vspace{-0.5em}
\end{figure}

As the radar signal reflected from a walking person is highly non-stationary, its characteristics are best revealed in a joint-variable representation, such as the time-frequency domain. For human motion analysis, the spectrogram is employed to show the signal's power distribution over time and frequency. For a discrete-time signal $\tilde{s}(n)$, the spectrogram is given by the squared magnitude of the STFT \cite{Opp99}
\begin{equation}
\mathrm{S}(n,k) = \left| \sum_{m=0}^{M-1} w(m) \tilde{s}(n+m) \exp{\left(-j 2 \pi \frac{mk}{K}\right)}\right|^2, 
\label{eq:spectrogram}
\end{equation}
for $n = 0, \dots, N-1$, where $M$ is the length of the smoothing window $w(\cdot)$, $k$ is the discrete frequency index with $k = 0, \dots, K-1$, and $M, K \in \mathbb{N}$.Spectrograms for different walking styles and directions relative to the sensor are shown in Fig.~\ref{fig:specs}. 
We pointed out that normal stride signatures when moving toward the radar system are different compared to those when the radar has a back view on the target (compare Figs.~\ref{fig:specs}\subref{NW} and~\ref{fig:specs}\subref{NWa}) \cite{Sei17}. The salient micro-Doppler feature of a normal stride away from the radar is a spike, i.e., an impulsive-like behavior in the TFR. This characteristic has been overlooked in other works and was not reported in any experimental as well as simulated micro-Doppler signatures. The latter includes, for example, the approach of using a Microsoft Kinect sensor to estimate the human posture via 20 points on the skeleton \cite{Ero15}. Besides the restriction that the entire body needs to be in the field of view of the Kinect sensor, the number of discrete sensing points along the human body is not sufficient to capture fine details in human locomotion. As a result, strides toward the sensor appear similar to those away from it. On the other hand, using radar, we can clearly identify deviations from a normal stride, e.g., when one of the knees is not fully bent (see Figs.~\ref{fig:specs}\subref{L1} and~\ref{fig:specs}\subref{L1a}) or a cane is used (see Figs.~\ref{fig:specs}\subref{CW} and~\ref{fig:specs}\subref{CWa}).

From the spectrogram, two other signals can be generated: the envelope signal of the micro-Doppler signatures and the short-time energy signal of the radar return. Both signals are extracted from the noise-reduced spectrogram, where an adaptive thresholding technique is used to suppress the background noise in the TFR \cite{Kim09}. To extract the envelope signal an energy-based thresholding algorithm is applied \cite{Ero16}. It captures the time-varying maximal Doppler shifts throughout a gait cycle, irrespective of whether it corresponds to a leg or cane motion, and thus converts the time-frequency signal into a train of pulses along the time variable. In contrast, the short-time energy signal accounts for the inherent pattern of individual micro-Doppler signatures by summing over $K'$ Doppler bins in the spectrogram as
\begin{equation}\label{eq:meanEnergy}
E(n) = \frac{1}{K'} \sum_{k=0}^{K'-1} \tilde{\mathrm{S}}(n,k), \quad n = 0, \dots, N-1,
\end{equation}
where $\tilde{\mathrm{S}}$ is the noise-reduced spectrogram, and $K' < K$ is the number of relevant frequency bins corresponding to micro-Doppler shifts excluding the torso's signature.

\subsection{Cadence-Velocity Diagram}
\begin{figure}[!t]
	\vspace{-0.8em}
	\centering{
		\subfloat[Normal walk]{\includegraphics[clip, trim= 0 0 20 8,width=0.5\columnwidth]{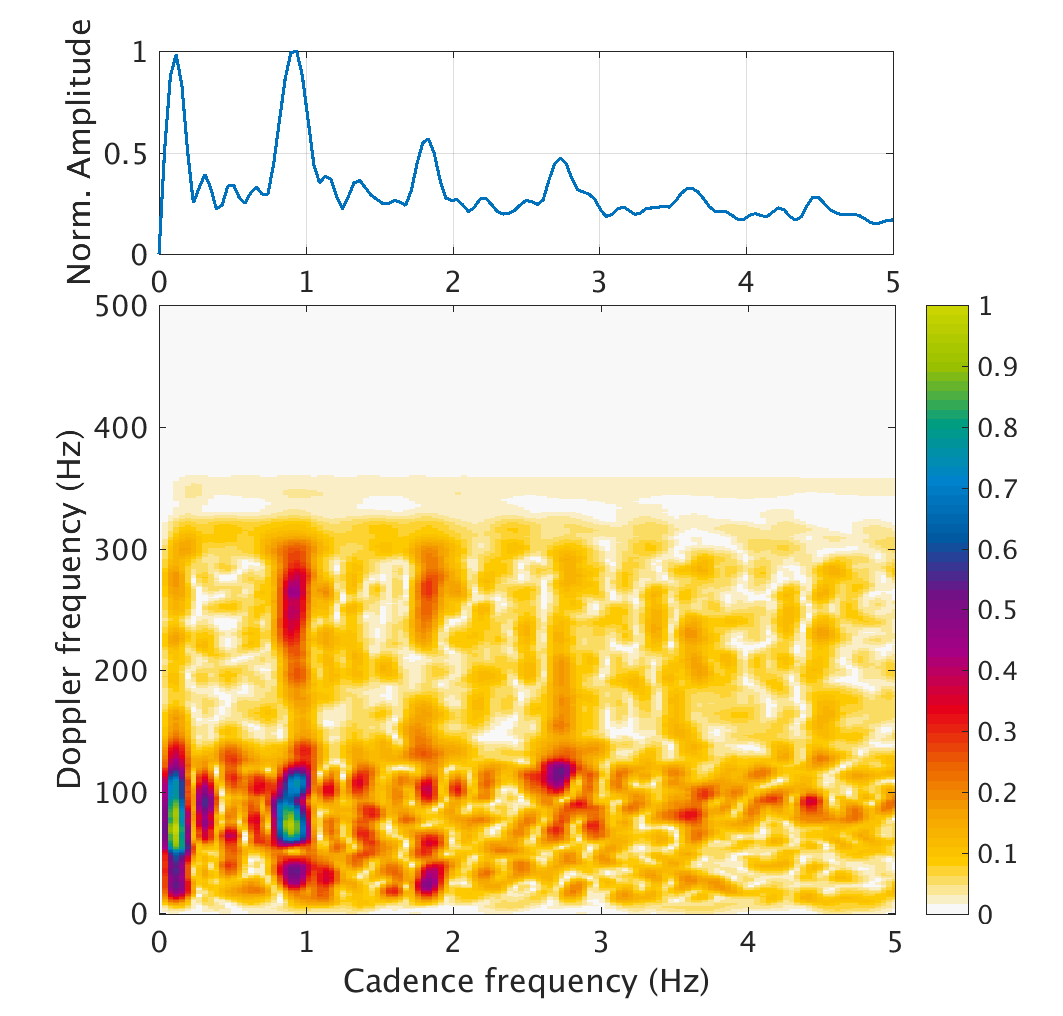}%
			\label{CVDNW}}
		\subfloat[Normal walk]{\includegraphics[clip, trim= 0 0 20 8,width=0.5\columnwidth]{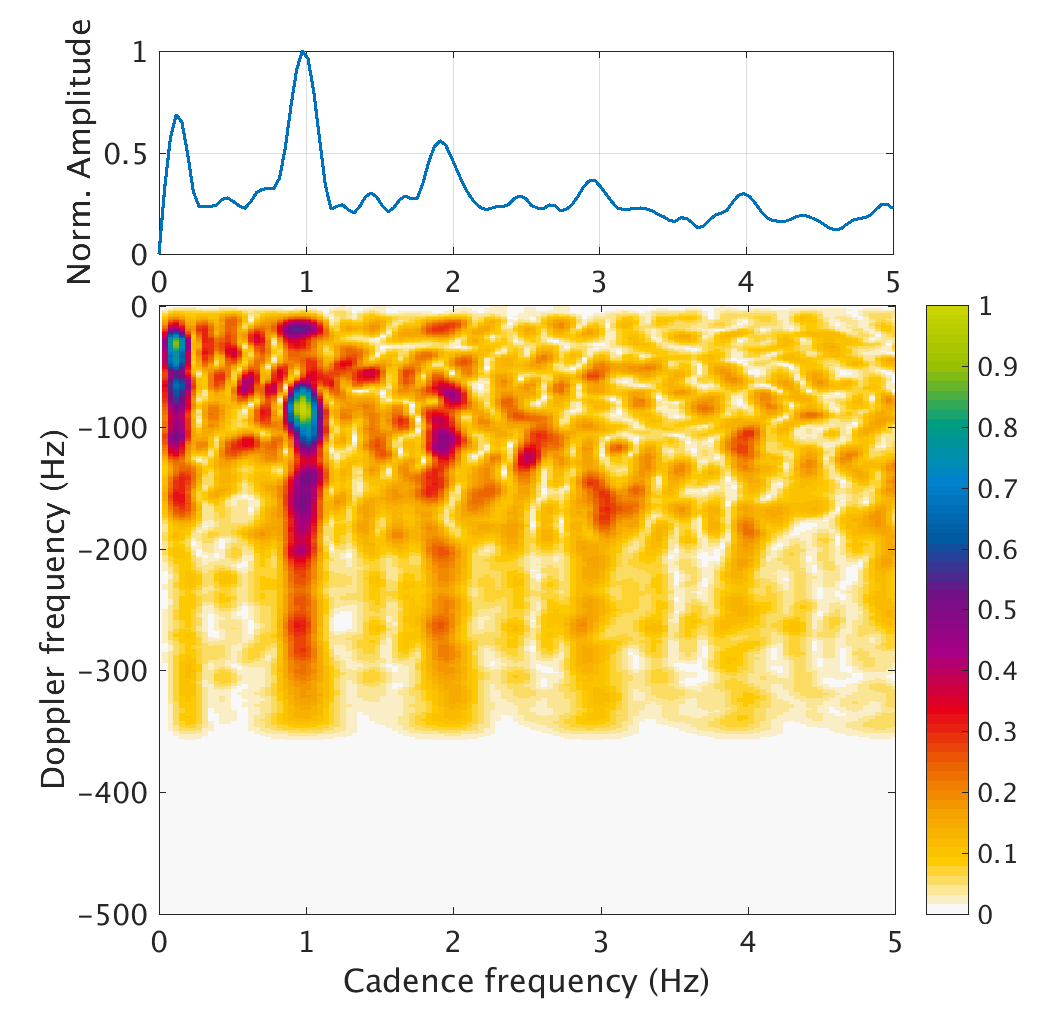}%
			\label{CVDNWa}}\vspace{-0.7em}}

	\centering{
		\subfloat[Limping with one leg]{\includegraphics[clip, trim= 0 0 20 8,width=0.5\columnwidth]{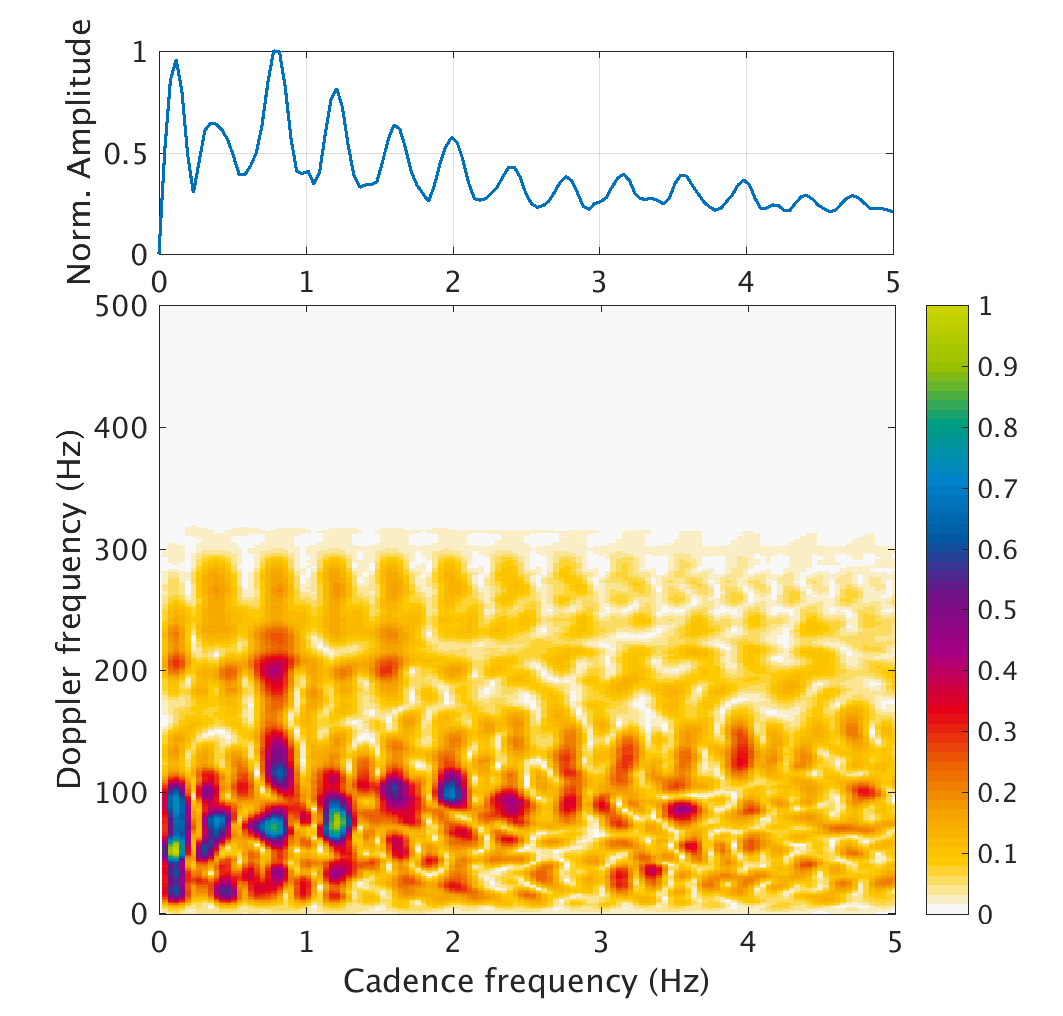}%
			\label{CVDL1}}
		\subfloat[Limping with one leg]{\includegraphics[clip, trim= 0 0 20 8,width=0.5\columnwidth]{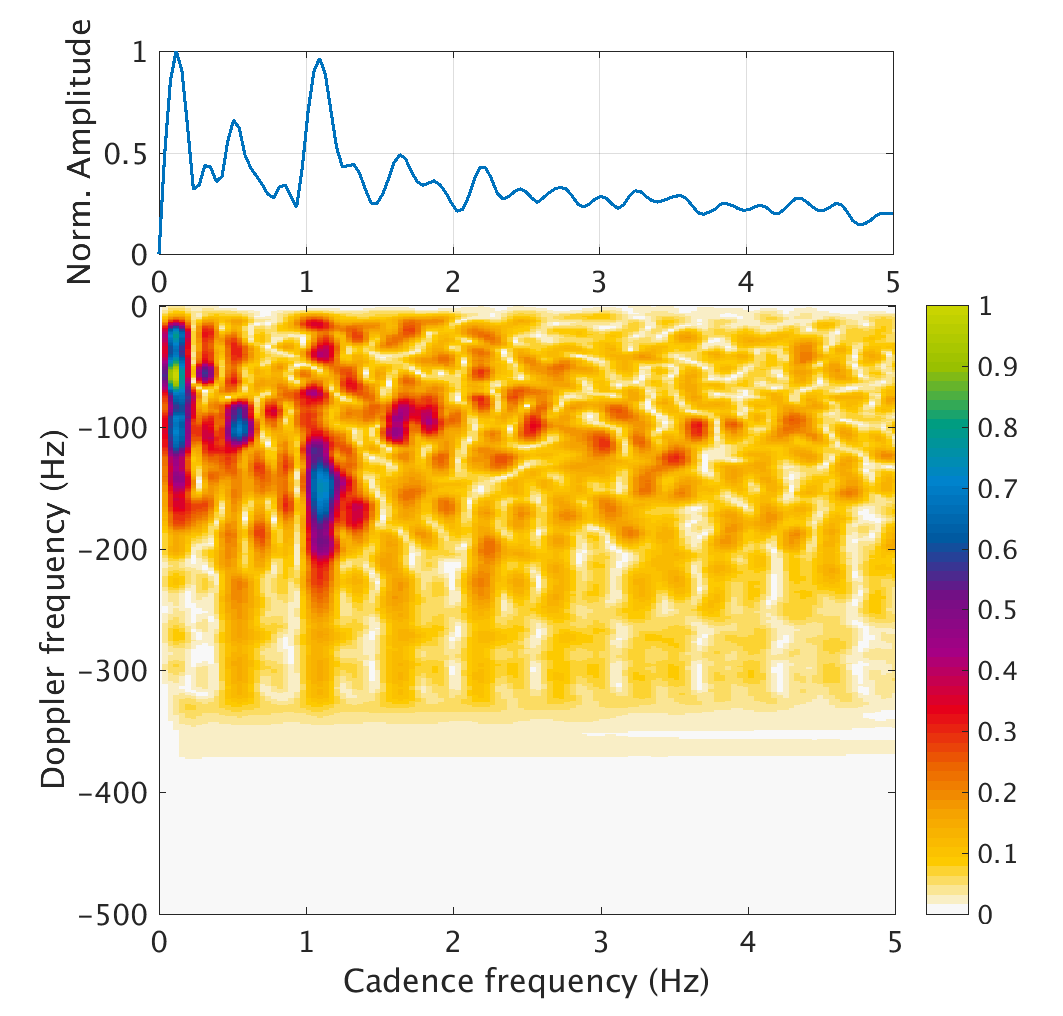}%
			\label{CVDL1a}}\vspace{-0.7em}}

	\centering{
		\subfloat[Walking with a cane]{\includegraphics[clip, trim= 0 0 20 8,width=0.5\columnwidth]{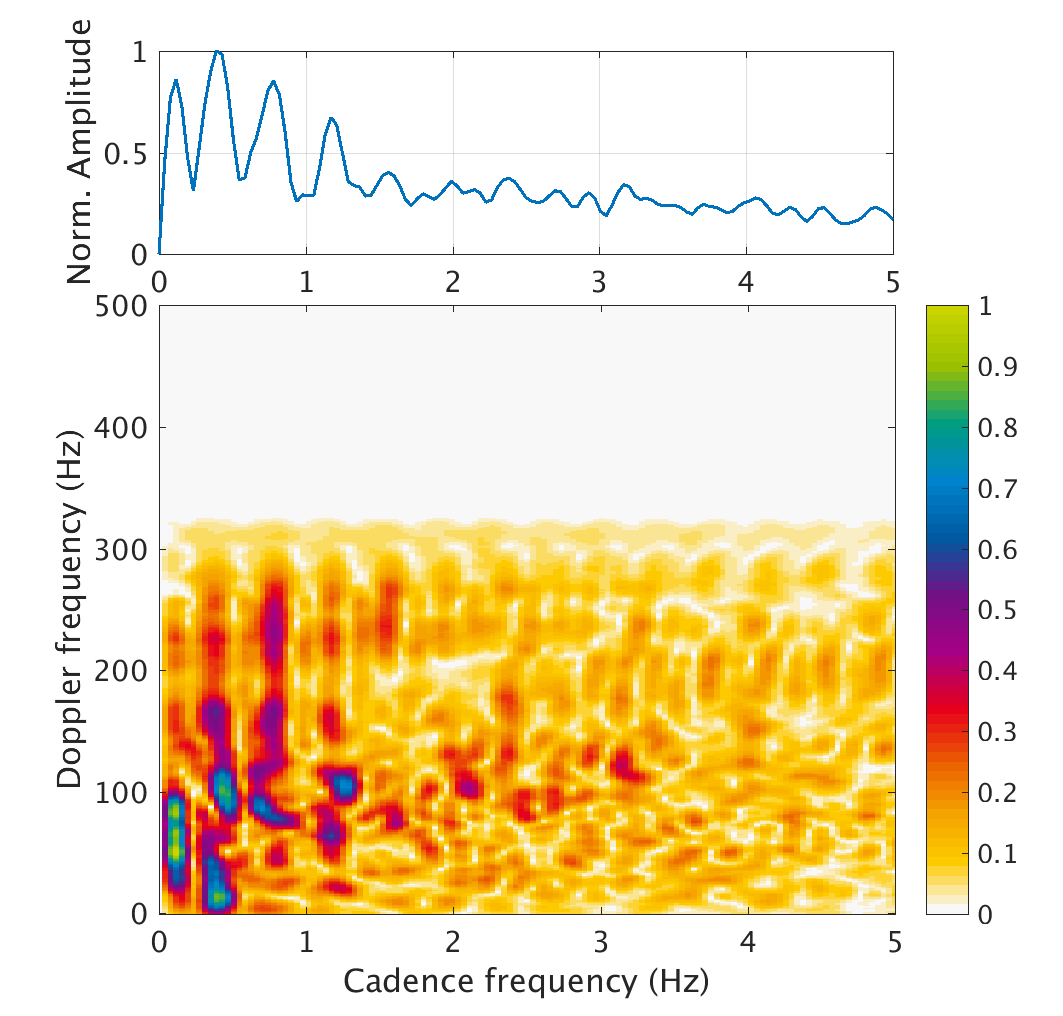}%
			\label{CVDCW}}
		\subfloat[Walking with a cane]{\includegraphics[clip, trim= 0 0 20 8,width=0.5\columnwidth]{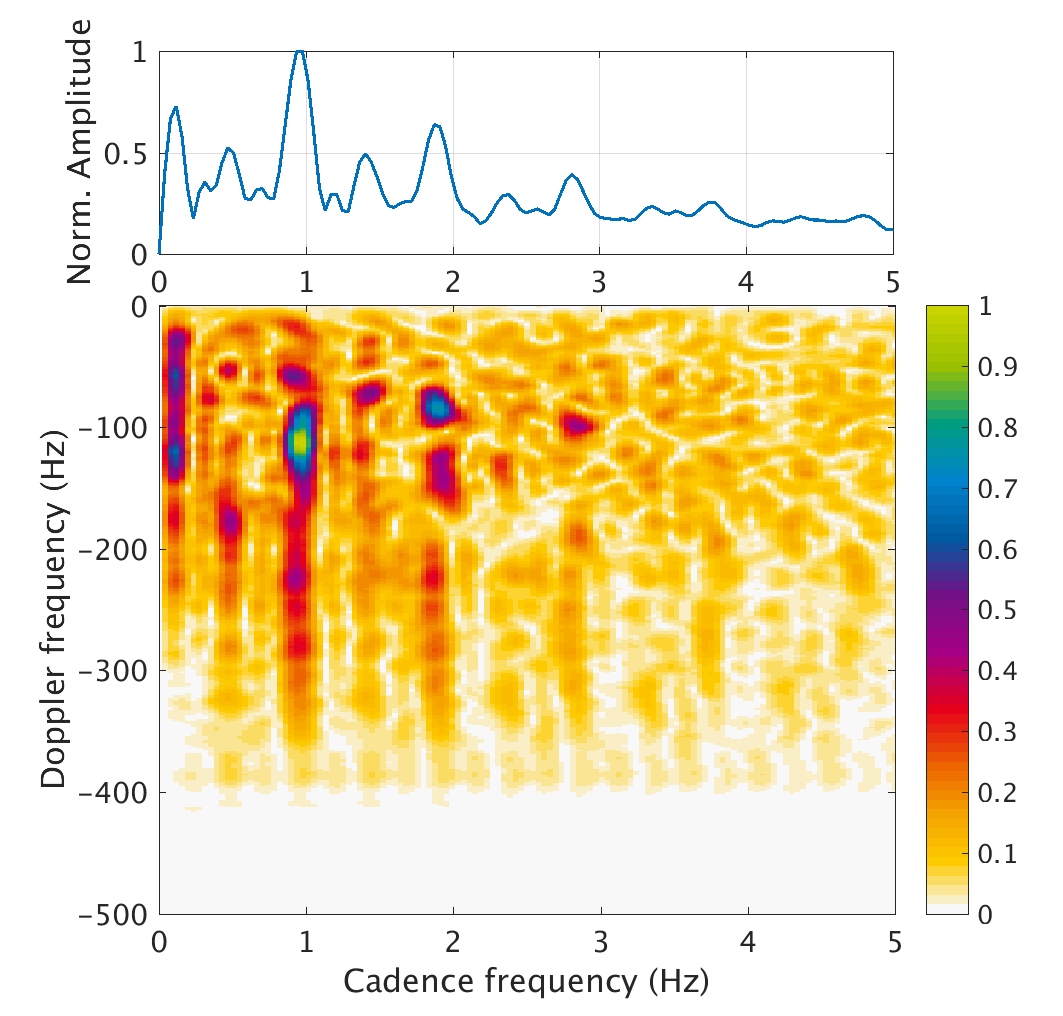}%
			\label{CVDCWa}}\vspace{-0.7em}}

	\centering{
		\subfloat[Walking with a cane out of sync]{\includegraphics[clip, trim= 0 0 20 8,width=0.5\columnwidth]{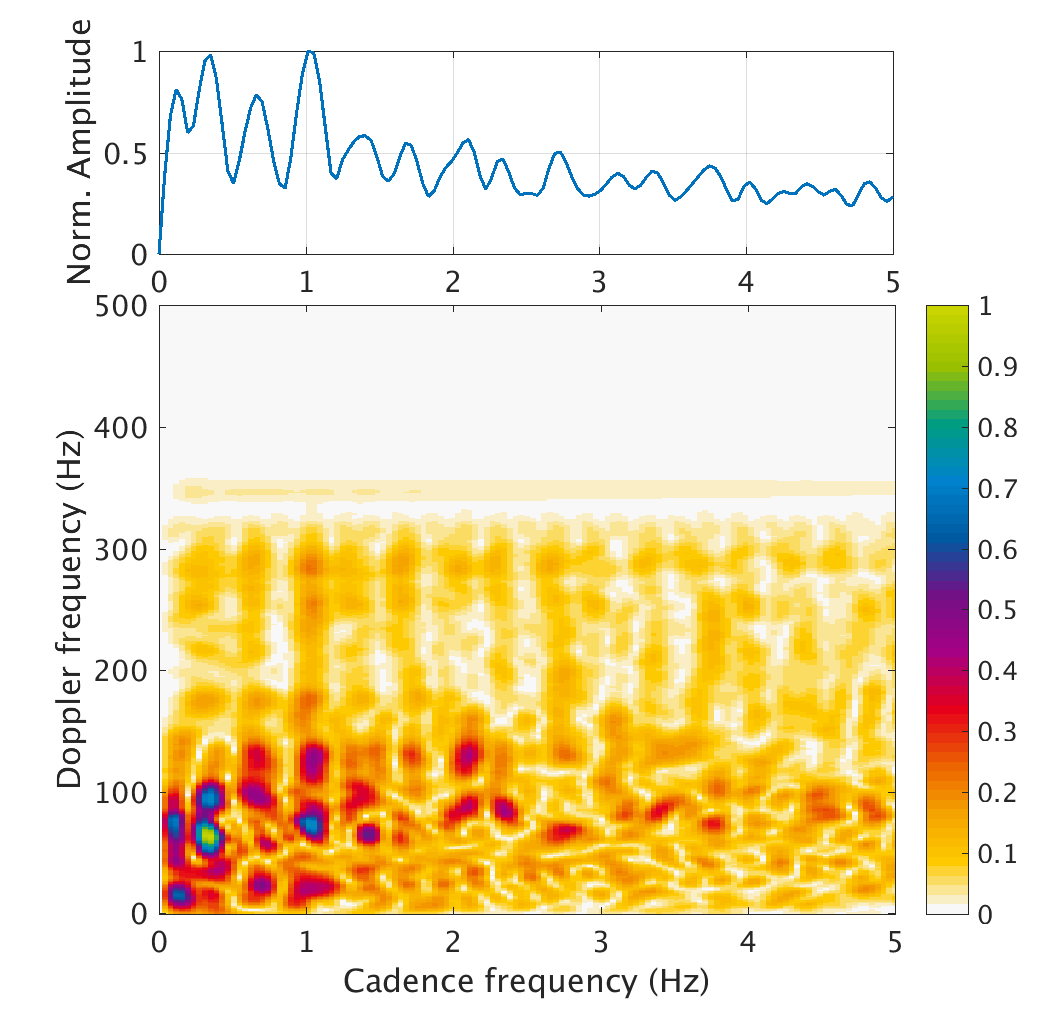}%
			\label{CVDCWoos}}
		\subfloat[Walking with a cane out of sync]{\includegraphics[clip, trim= 0 0 20 8,width=0.5\columnwidth]{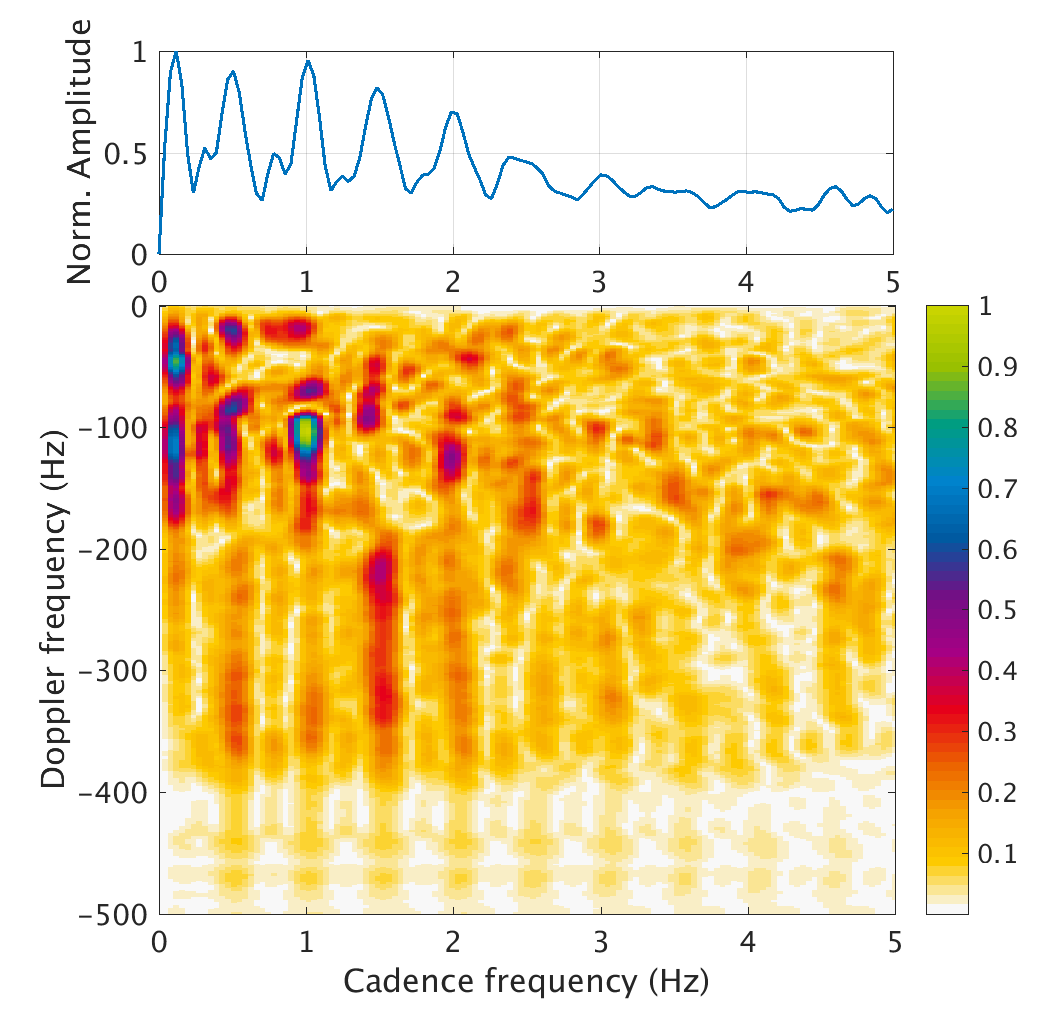}%
			\label{CVDCWoosa}}}

	\caption{Examples of CVDs and mean cadence spectra for walking toward (left) and away from (right) the radar system. The corresponding spectrograms are given in Fig.~\ref{fig:specs}.}
	\label{fig:cvds}\vspace{-0.5em}
\end{figure}

In order to analyze the cyclic nature and inherent periodicities of gait motions, we generate another joint-variable representation that depicts the periodic pattern of certain micro-Doppler components over time. This representation is the CVD which is obtained by taking the FT of the spectrogram along each Doppler frequency bin as \cite{Cle15}
\begin{equation}
\mathrm{C}(\epsilon,k)  = \left| \sum_{n=0}^{N-1} \tilde{\mathrm{S}}(n,k\textbf{}) \exp{\left(-j 2 \pi \frac{n \epsilon}{L}\right)} \right|, 
\label{eq:CVD}
\end{equation}
where $\epsilon = 0, \cdots, L-1$ is the cadence frequency, $\tilde{\mathrm{S}}$ is the noise-reduced spectrogram, and $L \in \mathbb{N}$. Here, the velocity is directly proportional to the observed Doppler frequency as given by (\ref{eq:Doppler}). Prior to FT calculation, the mean is removed from each Doppler frequency bin to reduce the influence of stationary Doppler components. Note that, in contrast to TFRs, the CVD does not depend on the initial phase of the gait cycle, i.e., it is a time-invariant analysis method. Fig.~\ref{fig:cvds} shows the corresponding CVDs of the measurements depicted in Fig.~\ref{fig:specs}, along with the mean cadence spectra.

The mean cadence spectrum (mCS) depicts how often certain Doppler components appear throughout a gait, independent of the components' velocities. It is obtained by summing over all Doppler frequencies in the CVD, i.e., \cite{Bjoe15,Ric15}
\begin{equation}\label{eq:meanCS}
\bar{\zeta}(\epsilon) = \frac{1}{K} \sum_{k=0}^{K-1} \mathrm{C}(\epsilon,k), \quad \epsilon = 0, \dots, L-1.
\end{equation}
The highest peak of the mCS typically represents the stride rate. For example, Fig.~\ref{fig:cvds}\subref{CVDNW} reveals a stride rate of 0.9\,Hz, which is consistent with almost six strides in the 6\,s data measurement (see Fig.~\ref{fig:specs}\subref{NW}). The step rate is an important characteristic of a walk and belongs to the group of physical features, which are easily interpretable. Note that, in medical terminology of gait analysis, the cadence is defined as the number of steps per unit time, and thus serves as a measure of the step rate \cite{Lev12}. Here, however, we draw a distinction between the cadence frequency, as a measure of the periodicity of micro-Doppler signatures, and the stride rate, which describes the number of strides per second. Therefore, the repetition frequency of micro-Doppler signatures, hereafter, is referred to as $\fmd$, which does not necessarily relate to the stride rate in assisted or pathological gait. 

Similarly, we can find the mean Doppler spectrum by summing over all cadence frequencies for each Doppler frequency bin in the CVD, i.e.,
\begin{equation}\label{eq:meanDS}
\bar{\Gamma}(k) = \frac{1}{L} \sum_{\epsilon=0}^{L-1} \mathrm{C}(\epsilon,k), \quad k = 0, \dots, K-1.
\end{equation}
From the mean Doppler spectrum, features such as the average walking speed of a person or the minimal and maximal Doppler shifts, can be extracted \cite{Ric15,Bjoe12}.
\section{Feature Extraction and Experimental Methods}\label{sec:featextraction}

\subsection{Experimental Data and Setup}\label{sec:exp_setup}
The experimental radar data were recorded in an office environment (no absorbers) at Technische Universit\"{a}t Darmstadt, Germany. An ultra-wideband radar \cite{Anc} was used in continuous-wave mode with a carrier frequency of 24\,GHz. The antenna feed point of the radar was placed at 1.15\,m above the floor, which represents a nominal hip height. Ten volunteers (age: 23.8 $\pm$ 2.6, 8 men, 2 women) were asked to walk slowly and without arm swinging toward and away from the radar system in an $0\degree$ angle relative to the radar LOS, between approximately 4.5\,m and 1\,m from the antenna feed point. All participants provided informed consent. Note that this setup would be a practice when radar is used as a diagnostic tool for analysis of gait abnormalities, where weak micro-Doppler signatures due to large aspect angles can easily be avoided. In total, the data set contains 1000 measurements, i.e., 100 measurements per person. Five different walking styles are considered: normal walking (NW), limping with one (L1) or both legs (L2), walking with a cane in sync with one leg (CW) and out of sync with any leg (CW/oos). Here, a limping leg is simulated by a knee that cannot be bent such that the stride motion is performed in a semicircular manner. In the case of limping with both legs neither of the knees can be bent. The number of samples per class and walking direction are equal among the test subjects. 

Using the same experimental setup, radar data of four additional subjects (4 women) with diagnosed gait disorders were collected at Villanova University, USA. This test data set contains 13, 20, 28 (16 thereof with a cane in sync with one leg), and 26 measurements for person A, B, C, and D, respectively, i.e., 87 measurements in total.

\subsection{Physical Features based on Sum-of-Harmonics Analysis}\label{subsec:phyfeatures}
Gait characteristics manifest themselves differently depending on the data representation and the transforms adopted. For feature extractions, we consider both spectrogram and CVD. Whereas the former depicts the Doppler and micro-Doppler signatures which correspond to velocities and their time-varying natures, the latter accentuates periodicities and better describes the harmonic components of the limbs. 

Concerning the feature extraction mechanism, we note that physical gait features are inter-related, and depend on data pre-processing, e.g., noise reduction in the spectrogram, type of radar used, environment, and target characteristics \cite{Bjoe15,Gur15}. Since the features play an important role in classification problems, they should be chosen to be (i) relevant to the considered classification problem and (ii) accurately estimated or extracted from the micro-Doppler signature or its transforms \cite{Gur15}. Features of physical interpretations have been widely used for radar-based human activity recognition and include, but are not limited to \cite{Kim09,Ote05}: 
\begin{itemize}
	\item torso Doppler frequency,
	\item total Doppler bandwidth,
	\item offset of the total Doppler,
	\item Doppler bandwidth without micro-Doppler effects,
	\item normalized standard deviation of Doppler signal strength,
	\item period of the limb motion or stride rate,
	\item average radial velocity,
	\item stride length,
	\item radar cross-section of some moving body components (gait amplitude ratio).
\end{itemize}
However, most of the above features are not descriptive for distinguishing different walking styles. For example, the target's average radial velocity is expected to be similar for the different classes of gait considered. Hence, the offerings of these features in gait recognition need to be examined.

In order to estimate the average radial velocity of a person, referred to as base velocity $v_0$, the mean Doppler spectrum is calculated as given by (\ref{eq:meanDS}). This mean value can be smoothed to minimize the influence of noise by applying a moving average filter with a span corresponding to approximately 11\,Hz in Doppler frequency or 0.07\,m/s. Next, the maximum value of the mean Doppler spectrum is determined, and the corresponding Doppler frequency serves as an estimate of $v_0$ by using (\ref{eq:Doppler}).

An important characteristic of a person's walk is the gait periodicity, which corresponds to the stride rate for a normal walk. However, in the case of cane-assisted walks, the strides may not be periodic due to the additional cane movements (see Figs.~\ref{fig:specs}\subref{CWoos} and \ref{fig:specs}\subref{CWoosa}). Hence, we introduce the micro-Doppler repetition frequency $\fmd$, which captures the periodicity of the micro-Doppler signatures irrespective of being due to leg or cane movements. For extracting $\fmd$, the spectrogram, given by (\ref{eq:spectrogram}), is used and the upper and lower envelopes of the micro-Doppler signatures are extracted for toward and away from radar motions, respectively, by applying an energy-based thresholding technique \cite{Ero16}. Since a swinging foot or a cane motion assumes the highest Doppler shifts in gait motions, they lead to maxima in the absolute value of the envelope signal. Taking the FT of the envelope signal, we find $\fmd$ by determining the frequency with the maximal amplitude.

Next, we consider the maximal observed Doppler shift in the measurement as a feature. This is motivated by the observation that a limping leg has a lower radial velocity, and thus causes smaller Doppler shifts. The maximal Doppler shift $\fdmax$ is estimated by use of the maximal values of the envelope signal. Here, the mean of the highest 10\% of the maximal Doppler shifts is used to be less sensitive to variations between different micro-Doppler stride signatures. 

\begin{figure}[!t]	
	\subfloat[]{
		\centering
		\includegraphics[clip, trim= 0 0 15 0,width=0.85\columnwidth]{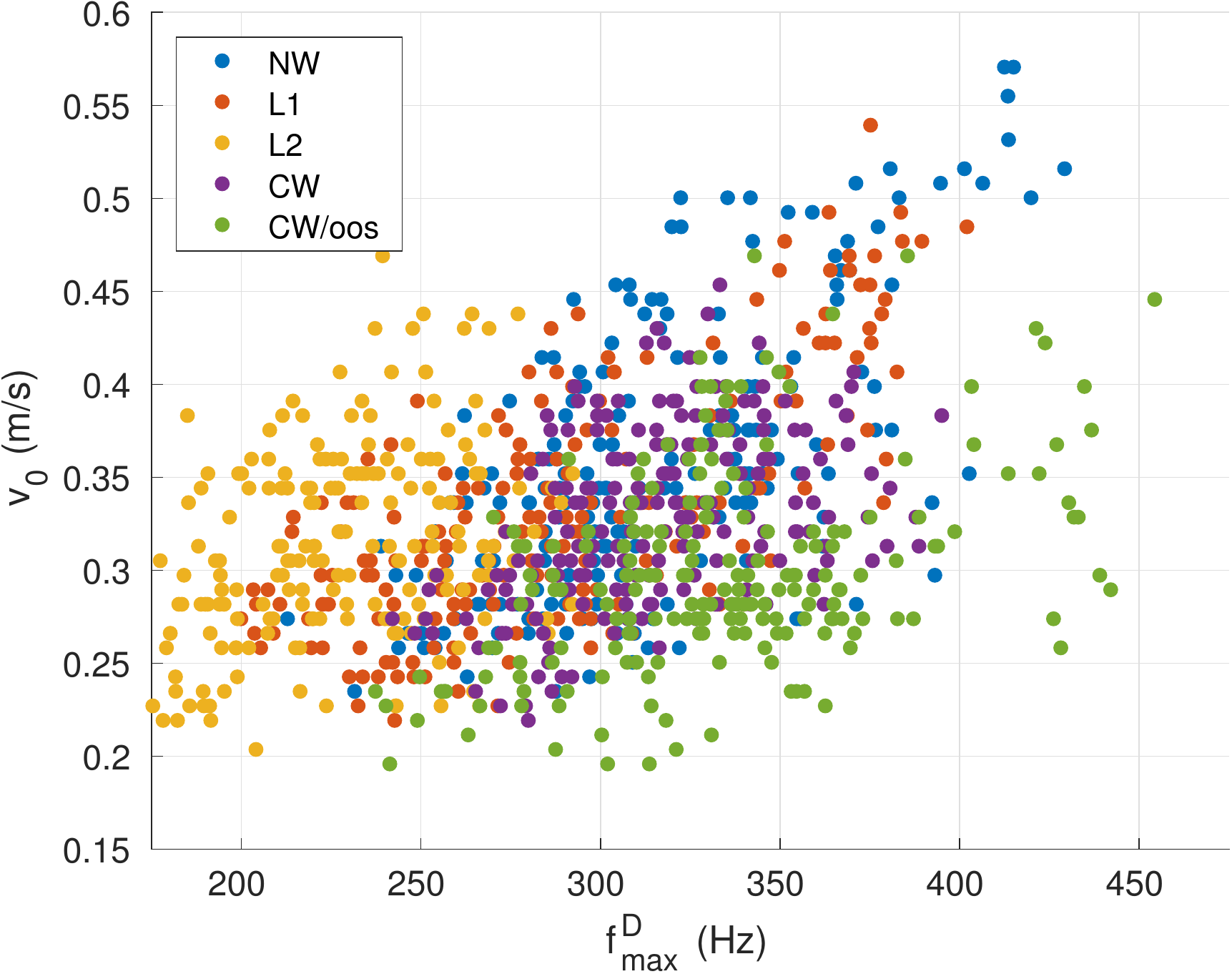}
		\label{fig:f0_vs_v0}} \\
	\subfloat[]{
		\centering
		\includegraphics[clip, trim= 0 0 15 0,width=0.85\columnwidth]{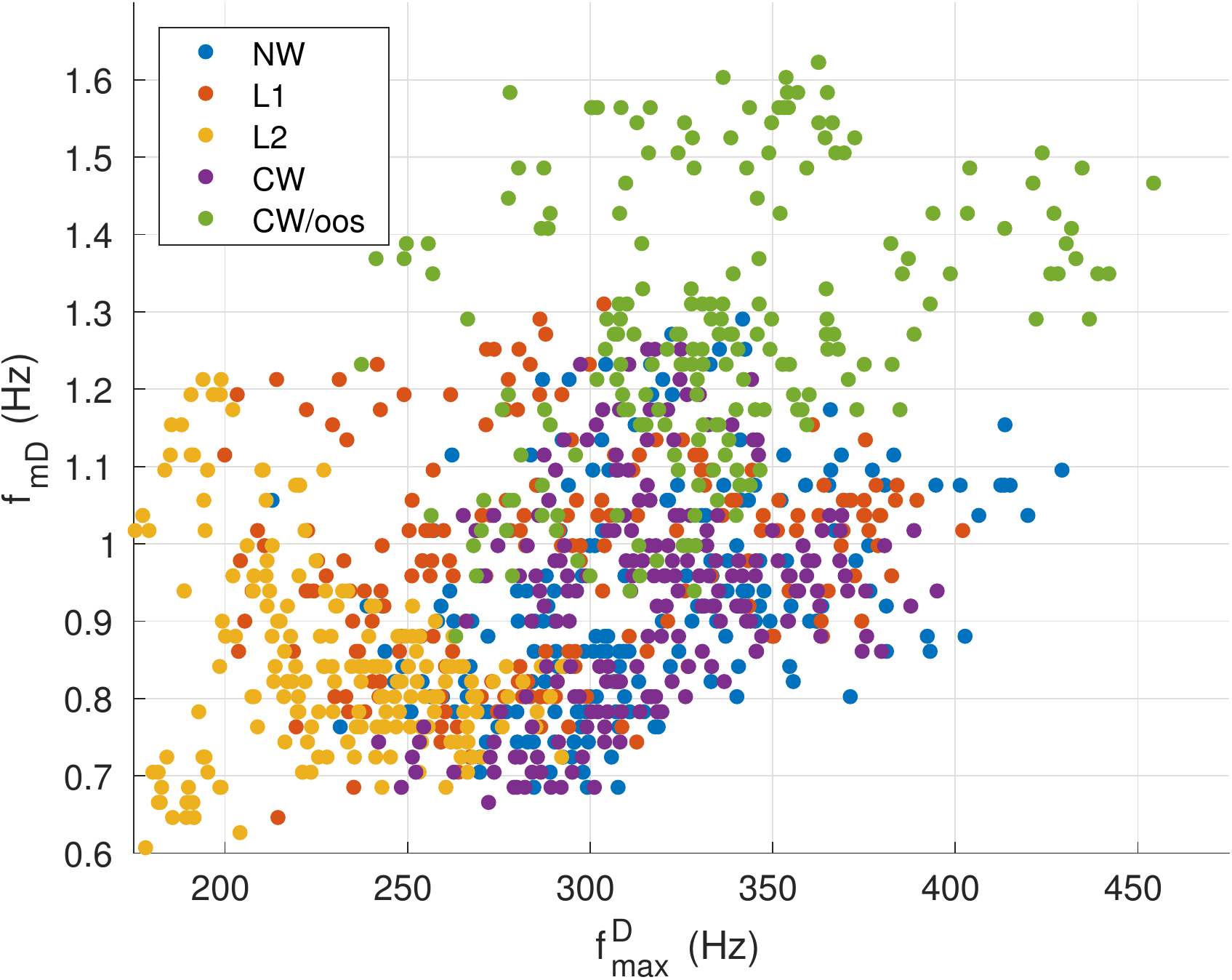}
		\label{fig:f0_vs_fdmax}}
	\caption{Scatter plots of physical features: (a) base velocity $v_0$ vs.~maximal Doppler shift $\fdmax$ and (b) micro-Doppler repetition frequency $\fmd$ vs.~$\fdmax$. \label{fig:scatterPhyFeat}}
	\vspace{-0.5em}
\end{figure}

In order to show relevance, Fig.~\ref{fig:scatterPhyFeat} shows scatter plots of the three described features, namely the base velocity $v_0$, the micro-Doppler repetition frequency $\fmd$, and the maximal Doppler shift $\fdmax$, for the five considered gait classes. From Fig.~\ref{fig:scatterPhyFeat}\subref{fig:f0_vs_v0}, it can be seen that the base velocity is not discriminative of the walks, because the walking speed of a person is not (necessarily) influenced by the use of an assistive walking device or gait impairments. However, the scatter plot in Fig.~\ref{fig:scatterPhyFeat}\subref{fig:f0_vs_fdmax} reveals that the micro-Doppler repetition frequency increases when walking with a cane. In particular, we remark that if the cane is moved out of sync, $\fmd$ becomes notably higher compared to the other four classes. Further, we note that limping with both legs has clearly the lowest maximal Doppler shift among the considered classes.

However, except for limping with both legs, the maximal Doppler shift does not help in discriminating between the remaining gait classes. In this respect, we note that some walking styles exhibit different maximum Doppler shifts per leg or cane motion. In particular, from Figs.~\ref{fig:specs}\subref{L1} and \ref{fig:specs}\subref{L1a}, we observe that limping with one leg leads to a characteristic pattern of alternating high and low maximal Doppler shifts. For capturing this oscillatory behavior, we find the peaks of the envelope signal and approximate the envelope's envelope using spline interpolation. In order to quantify the variation in maximal Doppler shifts, we proceed to calculate the coefficient of variation as $c_v = \frac{\sigma}{\mu}$, where $\sigma$ is the standard deviation and $\mu$ is the mean of the interpolated signal. The coefficient of variation is expected to be particularly high for limping with one leg and thus serves as an indicator of abnormality.

Further, we observe that the gait classes of normal walking (NW), limping with one leg (L1) and walking with a cane (CW) are not well separated in the feature space spanned by $\fmd$ and $\fdmax$ depicted in Fig.~\ref{fig:scatterPhyFeat}(b). That is, the micro-Doppler repetition frequency $\fmd$, by itself, does not capture the underlying regularity or irregularity of the walk. For this, we calculate the gait harmonic frequency ratio as \cite{Alz14}
\begin{equation}
\beta = \frac{f_0}{\fmd},
\label{eq:beta}
\end{equation}
where $f_0$ is the fundamental frequency (FF) of the gait. For the considered gait classes, we expect the values of $\beta$ to be:
\begin{itemize}
	\item $1$ for NW and L2 as each micro-Doppler stride signature assumes the same pattern,
	\item $1/2$ for L1 and CW as every other micro-Doppler signature appears the same,
	\item $1/3$ for CW/oos as a set of two strides and one cane movement constitutes one period.
\end{itemize}
In order to find $f_0$, we use the short-time energy signal defined in (\ref{eq:meanEnergy}) and model it as a sum-of-harmonics (SOH) \cite{Sei18}, i.e.,
\begin{equation}\label{eq:soh}
\begin{aligned}
x(n) &= \sum_{i=1}^{q} u_i \cos(2 \pi i f_0 n ) + v_i \sin(2 \pi i f_0 n ) \\
     &= \sum_{i=1}^{q} \alpha_i \cos(2 \pi i f_0 n + \phi_i),
\end{aligned} 
\end{equation}
where $f_0$ is the fundamental frequency in Hz, $q$ is the number of harmonics (NOH), and the harmonic amplitudes and phases are $\alpha_i$ and $\phi_i$, respectively. 
Here, we use the algorithm proposed in \cite{Whi03} to estimate the FF, the NOH, and harmonic amplitudes and phases. Assuming the energy signal $E(n)$ is composed of a SOH signal $x(n)$ and an additive white Gaussian noise component $u(n)$, i.e.,
\begin{equation}\label{eq:model}
E(n) = x(n) + u(n), \quad n = 0, \dots,N-1, 
\end{equation}
the parameters are then found by minimizing the squared-error between the data and the model, i.e.,
\begin{equation}\label{eq:LS}
\xi = \sum_{n = 0}^{N-1} \left| x(n) - E(n) \right|^2,
\end{equation}
and utilizing the nonlinear least squares (NLS) method for estimating $f_0$, which is augmented by a model order selection method for detecting $q$. For this, (\ref{eq:LS}) is jointly optimized over candidate FFs and candidate orders. We use $\fmd$ as an initial estimate of the FF. In a first step of the SOH algorithm, this estimate is refined by minimizing (\ref{eq:LS}) using an optimization technique. Next, candidate FFs are determined from the refined $f_0$ estimate for which the cost function defined by the NLS method is evaluated. At this point, we incorporate prior knowledge to limit computational costs in the joint-optimization for finding $f_0$ and $q$, and avoid overfitting. As described earlier, we expect $f_0$ to be $1/3 \cdot \fmd$, $1/2 \cdot \fmd$ or $1 \cdot \fmd$ given the initial FF estimate $\fmd$ is correct. Thus, the candidate FFs assume only the aforementioned values. Given the estimates for the FF and the NOH, the SOH model in (\ref{eq:soh}) is linear in the parameters $u_i$ and $v_i$. Thus, using the linear least-squares solution, the harmonic amplitudes $\alpha_i$ and phases $\phi_i$, $i=1,\dots,q$, can be computed in a closed-form as a function of $f_0$ and $q$. The estimated parameter vector is thus given by
\begin{equation}
\mathbf{h} = \left[ f_0~q~\alpha_1 \cdots \alpha_q ~\phi_1 \cdots \phi_q \right].
\label{eq:SOHparas}
\end{equation}

Given $f_0$, we proceed to calculate the gait harmonic frequency ratio $\beta$ using (\ref{eq:beta}). Table~\ref{tab:results_beta} shows the classification results using solely the $\beta$ feature for all considered walking styles. Clearly, $\beta$ is proving to be an important descriptive feature to characterize the analyzed walking patterns, as 70\%, 74\% and 90\% of the respective measurements show the expected gait harmonic frequency ratios $1$, $1/2$ and $1/3$, respectively. 

Based on the above results and the contributions of the various parameters, we form a physical feature vector as
\begin{equation}
\mathbf{z}^\text{phy} = \left[\fmd~\fdmax~c_v~\beta~\alpha_1 \cdots \alpha_{q_\text{max}} \right],
\label{eq:phyfeat}
\end{equation}
where again $\fmd$ is the micro-Doppler repetition frequency, $\fdmax$ is the maximal observed Doppler shift in the measurement, $c_v$ is the coefficient of variation of maximal micro-Doppler shifts, and $\beta$ is the gait harmonic frequency ratio. The harmonic amplitudes $\alpha_i$ relate to the height of the peaks in the mCS and help to discriminate different articulations of abnormality. Here, $q_\text{max}=5$ is the maximal order of the SOH model and $\alpha_i = 0~\forall~i > q$. Note that we do not include the base velocity $v_0$, as it was found not to be an appropriate discriminative feature for the motions considered.

\begin{table}[!t]
	\renewcommand{\arraystretch}{1.2}\setlength{\tabcolsep}{0.3em}
	\caption{Confusion matrices for classifying three different gait patterns using the gait harmonic feature $\beta$. Numbers are given in \%.}
		\label{tab:results_beta}
		\centering
		\subfloat[toward and away motions]{
			\begin{tabular}{ l L{1.1cm} L{1.1cm} L{1.1cm}}
				\toprule
				\textbf{True / Predicted } & NW, L2 & CW, L1 & CW/oos \\
				\cmidrule{1-4}
				NW, L2 		& $\bm{69.50}$ & $26.75$ & $3.75$ \\				
				CW, L1 		& $21.75$ & $\bm{73.75}$ & $4.50$ \\				
				CW/oos		& $5.00$  & $5.50$  & $\bm{89.5}$ \\
				\bottomrule
			\end{tabular} \label{tab:beta_both}}\\
		\subfloat[toward motions]{
			\begin{tabular}{ l L{1.1cm} L{1.1cm} L{1.1cm}}
				\toprule
				\textbf{True / Predicted } & NW, L2 & CW, L1 & CW/oos \\
				\cmidrule{1-4}
				NW, L2 		& $\bm{69.5}$ & $26.00$ & $4.50$ \\		
				CW, L1 	    & $9.50$ & $\bm{84.00}$ & $6.50$\\				
     			CW/oos		& $3.00$ & $5.00$ &  $\bm{92.00}$ \\
				\bottomrule
			\end{tabular} \label{tab:beta_toward}}\\
		\subfloat[away motions]{
			\begin{tabular}{ l L{1.1cm} L{1.1cm} L{1.1cm}}
				\toprule
				\textbf{True / Predicted } & NW, L2 & CW, L1 & CW/oos \\
				\cmidrule{1-4}
				NW, L2		& $\bm{69.50}$ & $27.50$ & $3.00$\\
				CW, L1  	& $34.00$ & $\bm{63.50}$ & $2.50$\\
				CW/oos		& $7.00$  & $6.00$ & $\bm{87.00}$\\
				\bottomrule
			\end{tabular} \label{tab:beta_away}}
		\vspace{-1em}
\end{table}

In this work, we desire to compare the classification performance using the above features with those used by recent works in this field, particularly \cite{Bjoe15} and \cite{Ric15}. We limit our comparison to the classification techniques that employ the cadence-velocity domain, as proposed in this paper. Bj{\"o}rklund \textit{et.~al} \cite{Bjoe15} used a 77\,GHz radar system to discriminate between the motions crawl, creep, walk, jog and run, which were performed by three test subjects. For classification they used features from the cadence-velocity domain and a support vector machine (SVM). By taking the average over all velocities in the CVD, they form the mCS, from which the three highest peaks are identified. At the corresponding cadence frequencies, denoted as $f_1$, $f_2$, and $f_3$, the velocity profiles $\Gamma_1$, $\Gamma_2$, and $\Gamma_3$ are extracted from the CVD, i.e., the energy distribution in the CVD for the given cadence frequencies as a function of the Doppler frequency. The velocity profiles are resampled to have 100 samples each. Further, the base velocity $v_0$ is extracted by finding the maximum in the mean Doppler spectrum. The corresponding feature vector is given by
\begin{equation}\label{eq:Bjoefull}
\mathbf{z}^\text{B1}  = [f_1~f_2~f_3~\Gamma_1~\Gamma_2~\Gamma_3~v_0],
\end{equation}
where $\Gamma_i$ denotes the resampled velocity profile at cadence frequency $f_i$, $i=1,\dots,3$, and $v_0$ is the base velocity.Further, they define a reduced feature vector as $\mathbf{z}^\text{B2} = [f_1~f_2~f_3~|v_0|]$, where the velocity profiles are not considered. Note that they drop the sign of the base velocity by taking the absolute value, i.e., they do take the direction of motion into account.

Ricci and Balleri \cite{Ric15} extracted features from the cadence-velocity domain for target recognition and identification. For that, four different subjects were walking on a treadmill in front of a 10\,GHz radar at constant speed. For discriminating the targets performing the same motion, the following features were extracted. From the mCS, they obtain an estimate of the person's stride rate. In our work, we resort to $\fmd$, which is more reliably estimated from the envelope of the micro-Doppler signatures. Next, a mean Doppler spectrum $\bar{\Gamma}_\text{mD}$ is formed around $\fmd$ by averaging over $\delta$ = 5 neighboring cadence frequencies corresponding to 0.825\,Hz cadence bandwidth. The second and third feature, $f^\mathrm{D}_\text{mD,min}$ and $f^\mathrm{D}_\text{mD,max}$, are found by determining the minimum and maximum Doppler frequencies in $\bar{\Gamma}_\text{mD}$ exceeding a predefined threshold $\gamma$ = 0.05. 
Thus, the first feature vector is defined as
\begin{equation} \label{eq:Ric1}
\mathbf{z}^\text{R1} = [\fmd~f^\text{D}_{\text{mD,min}}~f^\text{D}_{\text{mD,max}}].
\end{equation}
Second, the mean Doppler spectrum around $f_\text{mD}$ is used to define the feature vector $\mathbf{z}^\text{R2} = \bar{\Gamma}_{\text{mD}}.$

\subsection{Subspace Features}\label{subsec:pcafeatures}
In PCA, the intrinsic features of the considered walking styles are automatically learned and do not necessarily bear one-to-one correspondence to human motion kinematics \cite{Sei17a}. First, the input signals $\mathbf{C}$ are vectorized row-wise, i.e., $\mathbf{c} = \text{vec}\{\mathbf{C}^\mathrm{T}\} \in \mathbb{R}^{p \times 1}$, and stacked column-wise to form a data matrix $\mathbf{Y}$, such that
\begin{equation}
\mathbf{Y} = [\mathbf{c}_1~\mathbf{c}_2~\cdots~\mathbf{c}_d ] \in \mathbb{R}^{p \times d},
\end{equation} 
where $d$ is the number of training samples. The principal components are given by the eigenvectors of the covariance matrix.
There are various methods to compute the principal components \cite{Shl14}. We apply singular value decomposition (SVD) to decompose the data matrix such that $\mathbf{Y} = \mathbf{U} \mathbf{D} \mathbf{V}^\mathrm{T}$, where the columns of $\mathbf{U}$ and $\mathbf{V}$ are the left and right eigenvectors, respectively, and the diagonal entries of the diagonal matrix $\mathbf{D}$ are the singular values. The eigenvalues are related to the singular values by $\mathbf{\Lambda} = 1/(d-1) \mathbf{D}^2$ \cite{Koc13}.
The left eigenvector that has the largest eigenvalue, i.e., explains most of the variance in the data, is the first principal component. Using $\lambda$ principal components, which span a $\lambda$-dimensional subspace of the originally $d$-dimensional data space, each vectorized training and test image, $\mathbf{c}$, is projected onto that subspace by
\begin{equation}
\mathbf{p} = \tilde{\mathbf{U}}^\mathrm{T} \mathbf{c} \in \mathbb{R}^{\lambda \times 1},
\end{equation}
where $\tilde{\mathbf{U}} \in \mathbb{R}^{p \times \lambda}$ are the eigenvectors, or eigenimages, corresponding to the first $\lambda$ eigenvalues. The resulting projections $\mathbf{p}$ form the feature vector used for classification, i.e.,
\begin{equation}
\mathbf{z}^\text{PCA} = [p_1~p_2~\cdots~p_{\lambda}]^\mathrm{T}, \quad \lambda \leq d \in \mathbb{N}.
\label{eq:pcafeature}
\end{equation}

\subsection{Radar Data Representations}\label{ssec:inputsignals}
The different radar data representations and their dimension are listed in Table~\ref{tab:inputs}.

Using measurements of 6\,s duration, we calculate the spectrogram using (\ref{eq:spectrogram}), where a Hamming window of approximately 0.1\,s length is applied and the STFT is evaluated at K = 2048 frequency points. An excerpt of the spectrogram is used with Doppler components smaller than 500\,Hz as depicted in Fig.~\ref{fig:specs}, and its amplitude is normalized to the range of [0,1]. To further reduce the dimensionality of the spectrogram, it is sub-sampled in the time-domain by a factor of 20 and image binning is applied, where groups of 4 $\times$ 4 pixel are averaged. Thus, the spectrogram has 101 $\times $192 = 19392 entries. 

Next, the CVD is calculated according to (\ref{eq:CVD}), where zero-padding is used to obtain a cadence frequency resolution of approximately 0.04\,Hz. Again, the relevant part of the CVD images is extracted. In this regards, cadences up to 5\,Hz and Doppler frequencies from 0\,Hz to +500\,Hz and 0\,Hz to \mbox{-500\,Hz} for toward and away from radar motion measurements are considered, respectively, as shown in Fig.~\ref{fig:cvds}. Further, the resulting CVD image is downsampled in the Doppler domain to yield an image of dimensions 101 $\times$ 129 pixels, and is normalized to have values in the range of [0,1]. From this excerpt of the CVD, the mCS is obtained via (\ref{eq:meanCS}).

As we are particularly interested in the characteristic pattern in the CVD image, different stride rates and different maximal Doppler shifts among the measurements are compensated so as to align the CVD images. This is achieved by warping the CVD images along the cadence frequency and Doppler frequency axis using $\fmd$ and $\fdmax$, respectively. Afterwards, all CVDs assume $\fmd$ = 1\,Hz and $\fdmax$ = 500\,Hz. These CVDs are again considered up to 5\,Hz cadence frequency resulting in images of dimension 101 $\times$ 129 pixels. Hereafter, these CVDs will be referred to as pre-processed CVDs. As for the raw CVD images, we can find the pre-processed mCS from the pre-processed CVDs using (\ref{eq:meanCS}).

Finally, one can alternatively utilize the time-domain signal itself, where the lower Doppler components due to the torso's motion are removed by high-pass filtering the signal with a cut-off frequency corresponding to 2$v_0$. Taking the FT of this high-pass filtered signal, we obtain a similar representation as the mCS with peaks at the fundamental cadence and its harmonics.

\begin{table}[!t]
	\renewcommand{\arraystretch}{1.3}
	\caption{Radar data representations and their dimensions.}
	\label{tab:inputs}
	\centering
	\begin{tabular}{l c c }
		\toprule
		Representation & Dimension & $p$ \\
		\cmidrule{1-3}
		Spectrogram			    					& $101 \times 192$ & $19392$\\
		Cadence-velocity diagram (CVD)  			& $101 \times 129$ & $13029$ \\
		Mean cadence spectrum (mCS)					& $1   \times 129$ & $129$ \\
		Pre-processed CVD 							& $101 \times 129$ & $13029$ \\
		Pre-processed mCS 							& $1   \times 129$ & $129$ \\
		FT of filtered time-domain signal 			& $1   \times 129$ & $129$ \\
		\bottomrule
	\end{tabular}
	\vspace{-0.5em}
\end{table}

\subsection{Methodology and Parameter Optimization}\label{ssec:method}
In order to assess the appropriateness of the proposed features to the intra gait motion classification problem at hand, a very simple classifier is considered, namely the nearest neighbor (NN) classifier. Final classification results are obtained utilizing 10-fold cross-validation, where stratified sampling is applied to preserve the original distribution of the classes in the training and test set. Where appropriate, we indicate the 95\% confidence interval for the estimated score. Besides classification accuracy, the false positive rate (FPR), false negative rate (FNR), and true positive rate (TPR) are used to assess a classifier's performance. Here, false positives are normal walks that are wrongly classified as abnormal or assisted, and false negatives refer to abnormal or assisted walks that are misclassified as normal walk. Finally, the proposed method is also evaluated using leave-one-subject-out cross-validation, where the test set contains data of one individual at a time.

\begin{figure}[!t]
	\centering
	\includegraphics[clip, trim= 0 0 20 18,width=0.9\columnwidth]{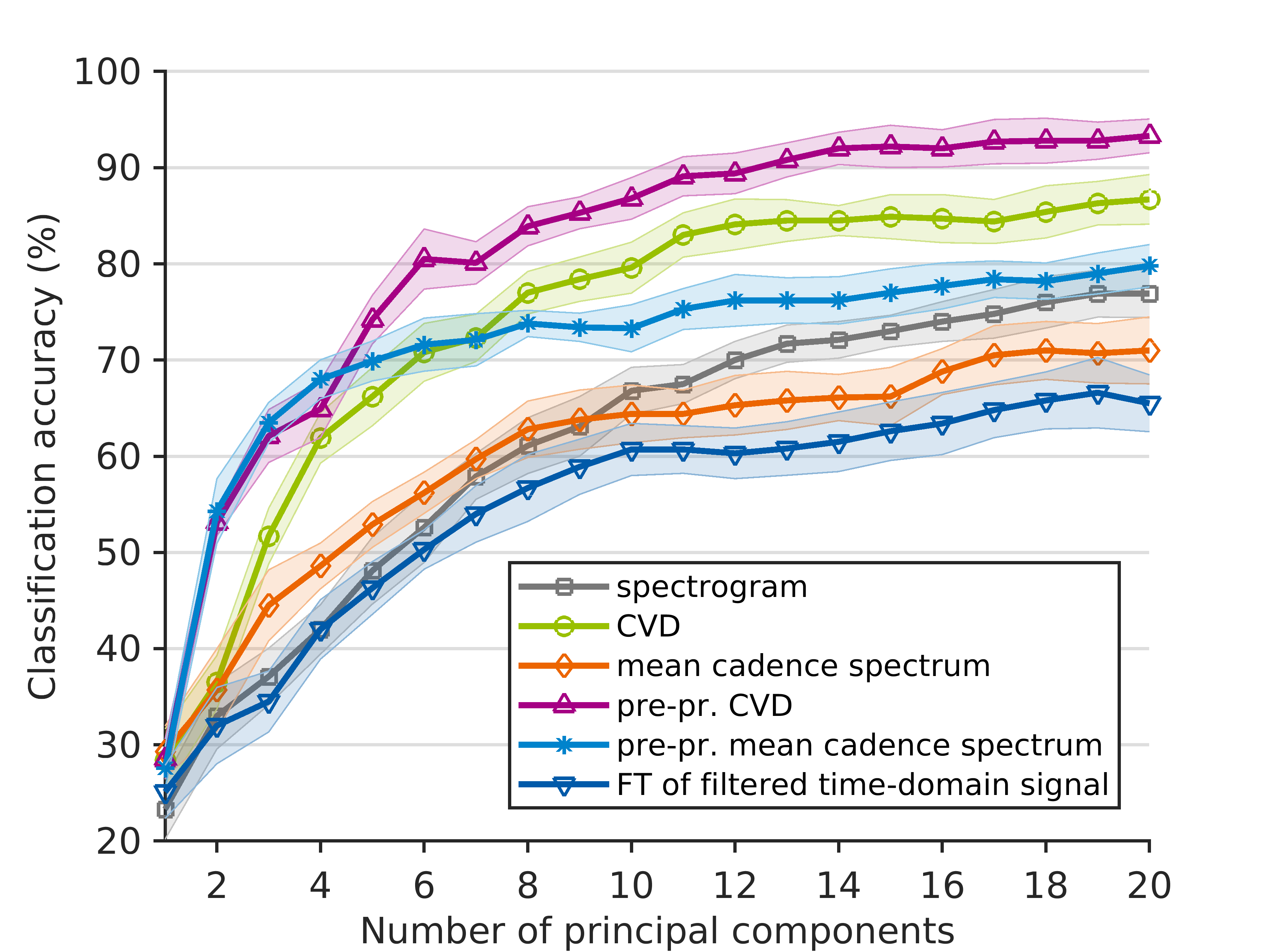}
	\caption{Average classification accuracy as a function of the number of principal components used for PCA-based feature extraction based on different radar data representations. The shaded areas depict the 95\% confidence intervals. \label{fig:parapet}}
	\vspace{-1em}
\end{figure}

Considering the different radar data representations listed in Table~\ref{tab:inputs}, we aim to find the optimal dimension of the PCA-based feature vector, as defined in (\ref{eq:pcafeature}), which depends on the number of principal components $\lambda$ used to span a subspace for data representation. In this work, we choose $\lambda$ such that the average classification accuracy is maximized. Toward this end, Fig.~\ref{fig:parapet} shows the achieved accuracies as a function of the number of principal components $\lambda$, where the classification accuracy does not significantly increase for $\lambda >$ 20 for any of the considered data representations. We find that the spectrogram is inferior to the other representations for extracting descriptive subspace-based features. One reason for the poor classification performance is that the spectrogram is a time-dependent representation of the human gait. The pre-processed CVD shows the highest classification rates. This indicates that the CVD contains key information relevant to classification which is lost when calculating the mCS. Based on these results, the pre-processed CVD is used as a reference in the follow-on comparison below.

\begin{figure}[!t]
	\centering
	\vspace{-0.5em}
	\includegraphics[clip, trim= 0 0 20 18,width=0.7\columnwidth]{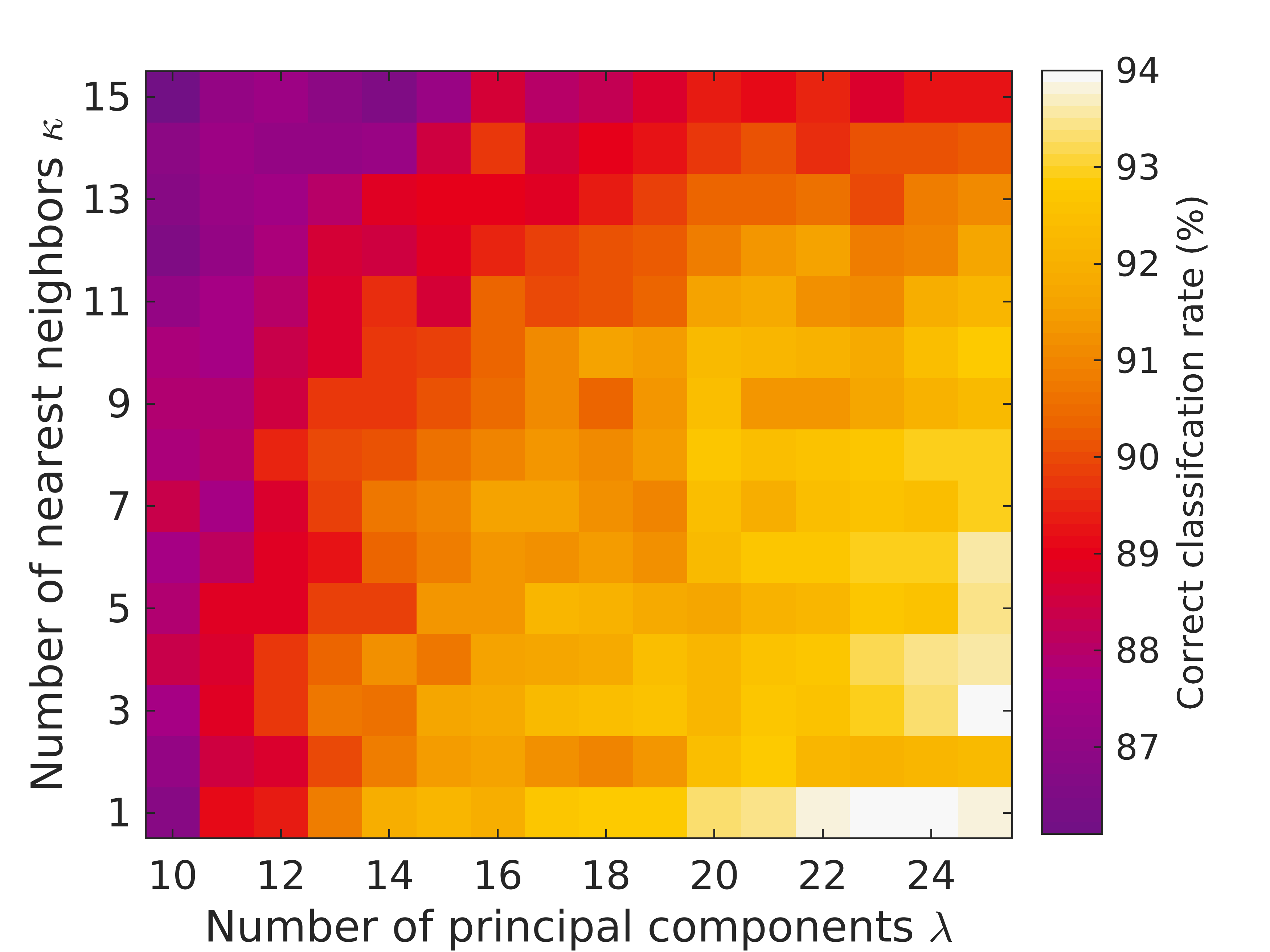}
	\caption{Average classification accuracy for different numbers of neighbors $\kappa$ for the classification process and different numbers of principal components $\lambda$ used for PCA-based feature extraction based on pre-processed CVDs. \label{fig:kappa_vs_lambda}}
	\vspace{-1em}
\end{figure}

In general, the NN classifier can be easily extended to the $\kappa$-NN classifier, which considers a number of $\kappa$ neighbors in the decision process. Similarly, the parameter $\kappa$ can be optimized such that the classifier achieves the highest average correct classification rate. Fig.~\ref{fig:kappa_vs_lambda} illustrates the joint optimization of the parameters $\lambda$ and $\kappa$ using pre-processed CVDs, where the color indicates the average correct classification rate. Note that we omitted the results for $\lambda <$ 10 for visual clarity, as the corresponding classification rates are significantly lower. We find that the highest accuracy is achieved by using the NN classifier ($\kappa$ = 1). Using more than $\lambda$ = 22 principal components in the PCA-based feature extraction process does not significantly increase the accuracy. Note that, larger values of $\kappa$ or $\lambda$ increases computation time.

\section{Experimental Results}\label{sec:results}

\begin{table}[!t]
	\renewcommand{\arraystretch}{1.2} \setlength{\tabcolsep}{0.25em}
	\caption{Confusion matrices using 10-fold cross-validation and different feature sets. Numbers are given in \%.} 	
	\centering
	\subfloat[physical features ($\mathbf{z}^{\text{phy}}$)]{
		\begin{tabular}{ l L{0.9cm} L{0.9cm} L{0.9cm} L{0.9cm} L{0.9cm}}
			\toprule
			\textbf{True / Predicted } & \multicolumn{1}{c}{NW} & \multicolumn{1}{c}{L1} & \multicolumn{1}{c}{L2} & \multicolumn{1}{c}{CW} & \multicolumn{1}{c}{CW/oos} \\
			\cmidrule{1-6}
			Normal walk (NW) 			& $\bm{82.0}$ 	& $1.5$ 		& $5.0$ 		& $9.0$ 		& $2.5$	\\
			
			Limping with one leg (L1) 	& $3.5$  		& $\bm{88.0}$ 	& $2.0$ 		& $3.5$ 		& $3.0$ \\
			
			Limping with both legs (L2) & $2.5$ 		& $1.5$ 		& $\bm{95.0}$ 	& $1.0$ 		& $0.5$ \\
			
			Cane - synchronized (CW) 	& $9.5$ 		& $5.5$ 		& $3.0$			& $\bm{80.0}$ 	& $2.0$ \\
			
			Cane - out of sync (CW/oos) & $2.0$ 		& $6.0$ 		& $2.0$  		& $2.0$ 		& $\bm{88.0}$\\
			\bottomrule
		\end{tabular} \label{tab:phy}}	
	
	\centering
	\subfloat[Bj\"orklund \textit{et al.} \cite{Bjoe15} ($\mathbf{z}^{\text{B1}}$)]{
		\begin{tabular}{ l L{0.9cm} L{0.9cm} L{0.9cm} L{0.9cm} L{0.9cm}}
			\toprule
			\textbf{True / Predicted } & \multicolumn{1}{c}{NW} & \multicolumn{1}{c}{L1} & \multicolumn{1}{c}{L2} & \multicolumn{1}{c}{CW} & \multicolumn{1}{c}{CW/oos} \\
			\cmidrule{1-6}
			Normal walk (NW) 			& $\bm{94.0}$ 	& $1.5$ 		& $1.5$ 		& $2.5$ 		& $0.5$	\\
			
			Limping with one leg (L1) 	& $8.5$  		& $\bm{78.0}$ 	& $5.5$ 		& $5.0$ 		& $3.0$ 	\\
			
			Limping with both legs (L2) & $1.5$ 		& $3.0$ 		& $\bm{90.5}$ 	& $2.5$ 		& $2.5$ 	\\
			
			Cane - synchronized (CW) 	& $13.0$ 		& $6.5$ 		& $5.5$			& $\bm{69.5}$ 	& $5.5$ \\
			
			Cane - out of sync (CW/oos) & $5.5$ 		& $7.0$ 		& $5.0$  		& $17.5$ 		& $\bm{65.0}$\\
			\bottomrule
		\end{tabular}\label{tab:Bjoe1}}	
	
	\centering
	\subfloat[Ricci and Balleri \cite{Ric15} ($\mathbf{z}^{\text{R1}}$)]{
		\begin{tabular}{ l L{0.9cm} L{0.9cm} L{0.9cm} L{0.9cm} L{0.9cm}}
			\toprule
			\textbf{True / Predicted } & \multicolumn{1}{c}{NW} & \multicolumn{1}{c}{L1} & \multicolumn{1}{c}{L2} & \multicolumn{1}{c}{CW} & \multicolumn{1}{c}{CW/oos} \\
			\cmidrule{1-6}
			Normal walk (NW) 			& $\bm{43.5}$ 	& $15.5$ 		& $11.0$ 		& $21.0$ 		& $9.0$	\\
			
			Limping with one leg (L1) 	& $14.5$ 		& $\bm{49.0}$ 	& $15.0$ 		& $15.0$ 		& $6.5$	\\
			
			Limping with both legs (L2) & $10.5$ 		& $10.0$ 		& $\bm{77.0}$ 	& $2.0$ 		& $0.5$	\\
			
			Cane - synchronized (CW) 	& $21.5$ 		& $11.0$ 		& $2.5$			& $\bm{57.5}$ 	& $7.5$ \\
			
			Cane - out of sync (CW/oos) & $8.0$ 		& $5.0$ 		& - 			& $10.5$ 		& $\bm{76.5}$\\
			\bottomrule
		\end{tabular}\label{tab:Ric1}}	
	
	\centering
	\subfloat[Ricci and Balleri \cite{Ric15} ($\mathbf{z}^{\text{R2}}$)]{
		\begin{tabular}{ l L{0.9cm} L{0.9cm} L{0.9cm} L{0.9cm} L{0.9cm}}
			\toprule
			\textbf{True / Predicted } & \multicolumn{1}{c}{NW} & \multicolumn{1}{c}{L1} & \multicolumn{1}{c}{L2} & \multicolumn{1}{c}{CW} & \multicolumn{1}{c}{CW/oos} \\
			\cmidrule{1-6}
			Normal walk (NW) 			& $\bm{78.5}$ 	& $5.0$ 		& $5.0$ 		& $9.5$ 		& $2.0$	\\
			
			Limping with one leg (L1) 	& $9.5$ 		& $\bm{73.0}$ 	& $9.0$ 		& $4.0$ 		& $4.5$	\\
			
			Limping with both legs (L2) & $4.5$ 		& $7.0$ 		& $\bm{83.5}$ 	& $4.0$ 		& $1.0$	\\
			
			Cane - synchronized (CW) 	& $9.0$ 		& $4.5$ 		& $3.5$ 		& $\bm{75.5}$ 	& $7.5$ \\
			
			Cane - out of sync (CW/oos) & $11.0$ 		& $5.0$ 		& $1.5$ 		& $18.0$ 		& $\bm{64.5}$\\
			\bottomrule
		\end{tabular}\label{tab:Ric2}}		

	\centering
	\subfloat[PCA-based features of CVDs ($\mathbf{z}^{\text{PCA}}$)]{
		\begin{tabular}{ l L{0.9cm} L{0.9cm} L{0.9cm} L{0.9cm} L{0.9cm}}
			\toprule
			\textbf{True / Predicted } & \multicolumn{1}{c}{NW} & \multicolumn{1}{c}{L1} & \multicolumn{1}{c}{L2} & \multicolumn{1}{c}{CW} & \multicolumn{1}{c}{CW/oos} \\
			\cmidrule{1-6}
			Normal walk (NW) 			& $\bm{93.5}$ 	& $1.0$ 		& $0.5$ 		& $4.5$ 		& $0.5$	\\
			
			Limping with one leg (L1) 	& - 			& $\bm{95.5}$ 	& - 			& $4.5$ 		& -	\\
			
			Limping with both legs (L2) & $1.5$ 		& $1.5$ 		& $\bm{93.0}$ 	& $4.0$ 		& -	\\
			
			Cane - synchronized (CW) 	& $6.5$ 		& $4.0$ 		& $0.5$ 		& $\bm{88.5}$ 	& $0.5$ \\
			
			Cane - out of sync (CW/oos) & $0.5$ 		& $0.5$  		& - 			& $0.5$ 		& $\bm{98.5}$\\
			\bottomrule
		\end{tabular}\label{tab:pca}}	
	\label{tab:class_results}
	\vspace{-1em}
\end{table}

\subsection{Physical Features}
Table \ref{tab:class_results}\subref{tab:phy} shows the classification results using the feature vector as defined in (\ref{eq:phyfeat}). The average correct classification rates assume 82.0\%, 88.0\%, 95.0\%, 80.0\% and 88.0\% for NW, L1, L2, CW and CW/oos, respectively. The overall accuracy is 86.6\%, with an FPR of 18.0\% and an FNR of 4.4\%. We note that normal walks (NW) are mostly confused with walking with a cane (CW), and vice versa. This is expected in the sense that the underlying motion of walking with a cane is a normal walk, where the cane's micro-Doppler signatures superimpose every other leg micro-Doppler signature.

Next, Table \ref{tab:class_results}\subref{tab:Bjoe1} presents the classification results for the feature vector $\mathbf{z}^{\text{B1}}$ used by Bj{\"o}rklund \textit{et. al}~\cite{Bjoe15}. For comparison, we use the same classifier as for the physical features, i.e., the NN classifier. Using the first feature vector, the overall correct classification rate assumes 79.4\%, with an FPR of 6.0\% and an FNR of 7.1\%. Despite the increased number of features, the average correct classification rate is lower compared to using physical features. Removing the velocity profiles from the feature vector, i.e., using $\mathbf{z}^{\text{B2}}$, the classification accuracy decreases to only 40.4\%, which shows that the cadence frequencies $f_1$, $f_2$ and $f_3$ along with the base velocity $v_0$ are not key in discriminating the considered gait classes.

Using the feature vectors $\mathbf{z}^{\text{R1}}$ and $\mathbf{z}^{\text{R2}}$ defined by Ricci and Balleri \cite{Ric15}, the results are given in Tables \ref{tab:class_results}\subref{tab:Ric1} and \subref{tab:Ric2}, respectively. In the first case, the parameters $\Delta m$ and $\gamma$ were optimized as to achieve the highest accuracy. The overall accuracy is found as 60.7\% with an FPR of 56.6\% and an FNR of 13.6\%. We again observe that normal walks (NW) are mostly confused with walking with a cane (CW) and vice versa. Using the mean Doppler spectrum around $\fmd$ as a feature, i.e., $\mathbf{z}^{\text{R2}}$, the accuracy is given by 75.0\%, where the FPR and FNR assume 21.5\% and 8.5\%, respectively. Again, we note that the mean Doppler spectrum comprises more information for classification of the considered motions than single Doppler or cadence frequencies.

Hence, we conclude that physical features, such as the base velocity or the micro-Doppler repetition frequency, are not suited to discriminate between the considered gait classes, i.e., to solve the intra motion category classification problem of gait recognition. However, signals obtained from the CVD, e.g., the mean Doppler spectrum, do hold discriminative characteristics that allow to distinguish between different walking styles.

\subsection{Subspace Features}
Table \ref{tab:class_results}\subref{tab:pca} shows the classification results utilizing PCA-based features of CVDs and the NN classifier. Here, $\lambda$ = 22 principal components are used. The overall accuracy assumes 93.8\%, where the FPR is 6.5\% and the FNR is 2.1\%. The highest classification rates are achieved for walking with a cane out of sync (98.5\%). The gait class CW shows the lowest classification rate (88.5\%) as this motion is again confused with normal walking, and vice versa. The results demonstrate the suitability of the CVD and the effectiveness of PCA for feature extraction. Even though we only consider one motion class and a single signal domain, the proposed method classifies the gaits with a high accuracy.

\begin{table}[!t]
	\renewcommand{\arraystretch}{1.2}  \setlength{\tabcolsep}{0.25em}
	\caption{Confusion matrix using leave-one-subject-out cross-validation and PCA-based features of CVDs. Numbers are given in \%.} 
	\label{tab:LOPOCV_PCA}
	\centering
	\begin{tabular}{ l L{0.9cm} L{0.9cm} L{0.9cm} L{0.9cm} L{0.9cm}}
		\toprule
		\textbf{True / Predicted } & \multicolumn{1}{c}{NW} & \multicolumn{1}{c}{L1} & \multicolumn{1}{c}{L2} & \multicolumn{1}{c}{CW} & \multicolumn{1}{c}{CW/oos} \\
		\cmidrule{1-6}
		Normal walk (NW) 			& $\bm{76.5}$ 	& $0.5$ 		& $2.5$			& $20.5$ 		& - \\
		
		Limping with one leg (L1) 	& $0.5$			& $\bm{83.5}$ 	& $0.5$ 		& $15.5$ 		& -	\\
		
		Limping with both legs (L2) & $9.0$ 		& $1.5$ 		& $\bm{83.0}$ 	& $6.5$ 		& -	\\
		
		Cane - synchronized (CW) 	& $17.0$ 		& $18.0$ 		& $3.0$ 		& $\bm{61.5}$ 	& $0.5$ \\
		
		Cane - out of sync (CW/oos) & - 	 		& $2.0$ 		& - 			& $0.5$			& $\bm{97.5}$\\
		\bottomrule
	\end{tabular}
\end{table}

Applying leave-one-subject-out cross-validation, the accuracy decreased to 80.4 $\pm$ 4.9\% (FPR 23.5 $\pm$ 15.5\%, FNR 6.6 $\pm$ 3.4\%). Here, $\lambda$ = 10 principal components are used and a $\kappa$-NN classifier is applied, where $\kappa$ = 24. The confusion matrix is shown in Table \ref{tab:LOPOCV_PCA}. Again, CW/oos shows the highest accuracy (97.5\%), while most of the confusion appears between normal (NW) and cane-assisted walks (CW). Since each person has its own walking style, the results are promising even though data of more persons are needed for generalization.

\begin{table}[!t]
	\renewcommand{\arraystretch}{1.2}
	\caption{Comparison of different gait recognition algorithms based on their classification accuracy (ACC), false positive rate (FPR) and false negative rate (FNR).}
	\label{tab:all_results}
	\centering
	\begin{tabular}{l r r r }
		\toprule
		Algorithm & \multicolumn{1}{c}{ACC (\%)}	& \multicolumn{1}{c}{FPR (\%)} & \multicolumn{1}{c}{FNR (\%)}	\\
		\cmidrule{1-4}
		Phy 			& $86.6 \pm 1.4$ 	& $18.0 \pm 4.7$ 	& $ 4.4 \pm 1.3$ 	\\	
		B1 \cite{Bjoe15}& $79.4 \pm 2.7$ 	& $ 6.0 \pm 3.5$ 	& $ 7.1 \pm 1.2$ 	\\	
		R1 \cite{Ric15} & $60.7 \pm 2.8$ 	& $56.6 \pm 5.9$ 	& $13.6 \pm 2.7$ 	\\
		R2 \cite{Ric15} & $75.0 \pm 3.1$ 	& $21.5 \pm 7.2$ 	& $ 8.5 \pm 1.6$ 	\\
		PCA 			& $93.8 \pm 1.5$ 	& $ 6.5 \pm 2.1$ 	& $ 2.1 \pm 0.7$ 	\\
		\bottomrule
	\end{tabular}
\end{table}

\subsection{Discussion}
We used radar measurements that contain a representative portion of the gait to classify five different walking styles, including abnormal and assisted gait. Table~\ref{tab:all_results} summarizes the results of all presented gait classification methods. The subspace feature extraction method utilizing PCA of CVD images achieves the highest correct classification rate (93.8\%), while the FPR (6.5\%) and the FNR (2.1\%) are kept low. Thus, we conclude that (i) subspace-based features are superior to physical features in classifying different gaits, and (ii) the CVD comprises more information on the gait than, e.g., the spectrogram. It is pointed out that the proposed method works reasonably good for all gait classes, despite of the relatively small number of 1000 measurements of ten different test subjects. This is certainly one benefit over popular deep learning approaches, which require a very large (training) data set and are computationally costly \cite{Sey18,Kim16}.

\begin{figure}[!t]
	\vspace{-1em}
	\centering{
		\subfloat[Person A]{\includegraphics[clip, trim= 0 0 20 18, width=0.5\columnwidth]{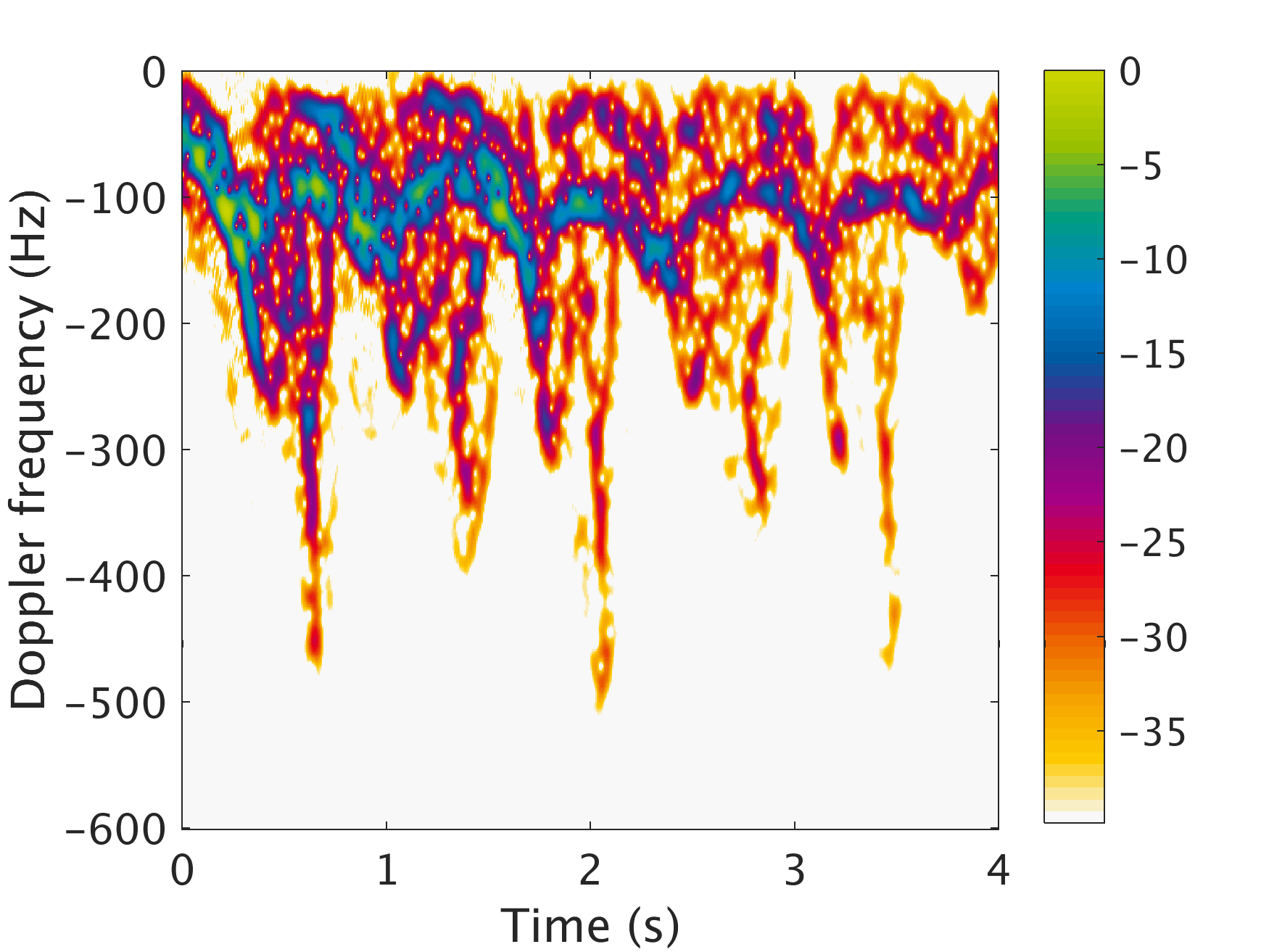}%
			\label{LA}}
		\subfloat[Person D]{\includegraphics[clip, trim= 0 0 20 18,width=0.5\columnwidth]{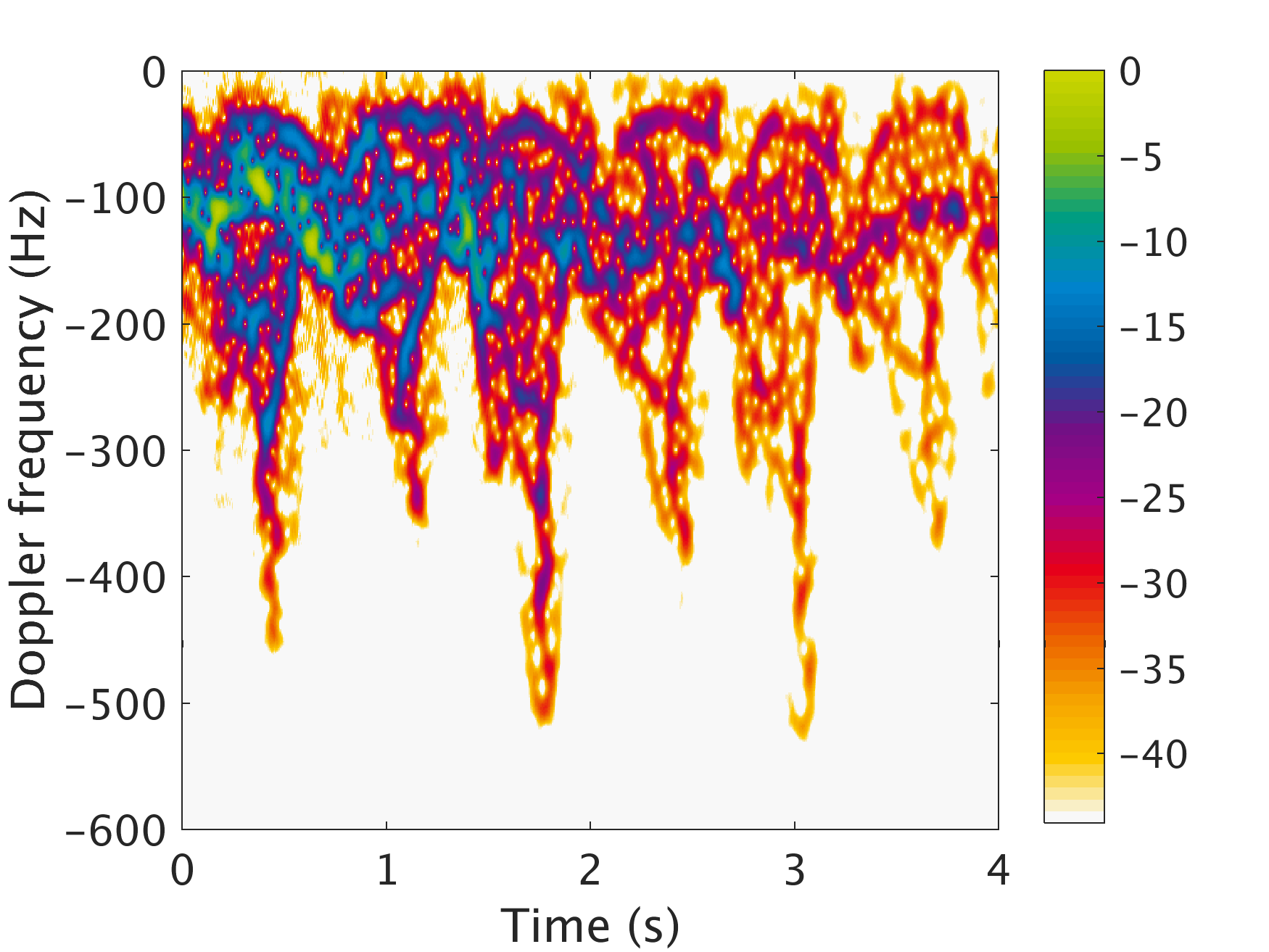}%
			\label{LD}}\vspace{-0.7em}}
	
	\centering{
		\subfloat[Person B]{\includegraphics[clip, trim= 0 0 20 18,width=0.5\columnwidth]{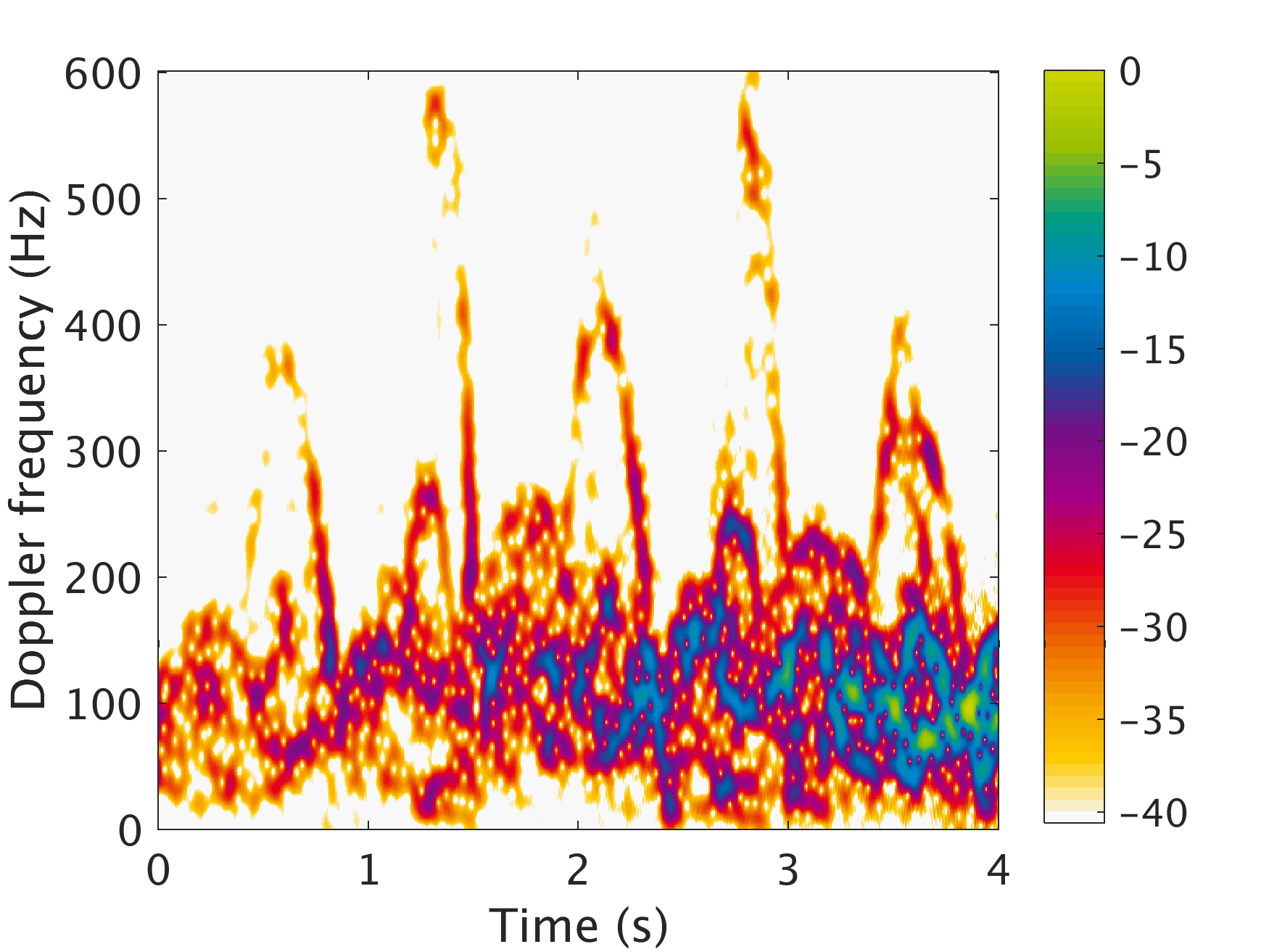}%
			\label{LB2}}
		\subfloat[Person B]{\includegraphics[clip, trim= 0 0 20 18,width=0.5\columnwidth]{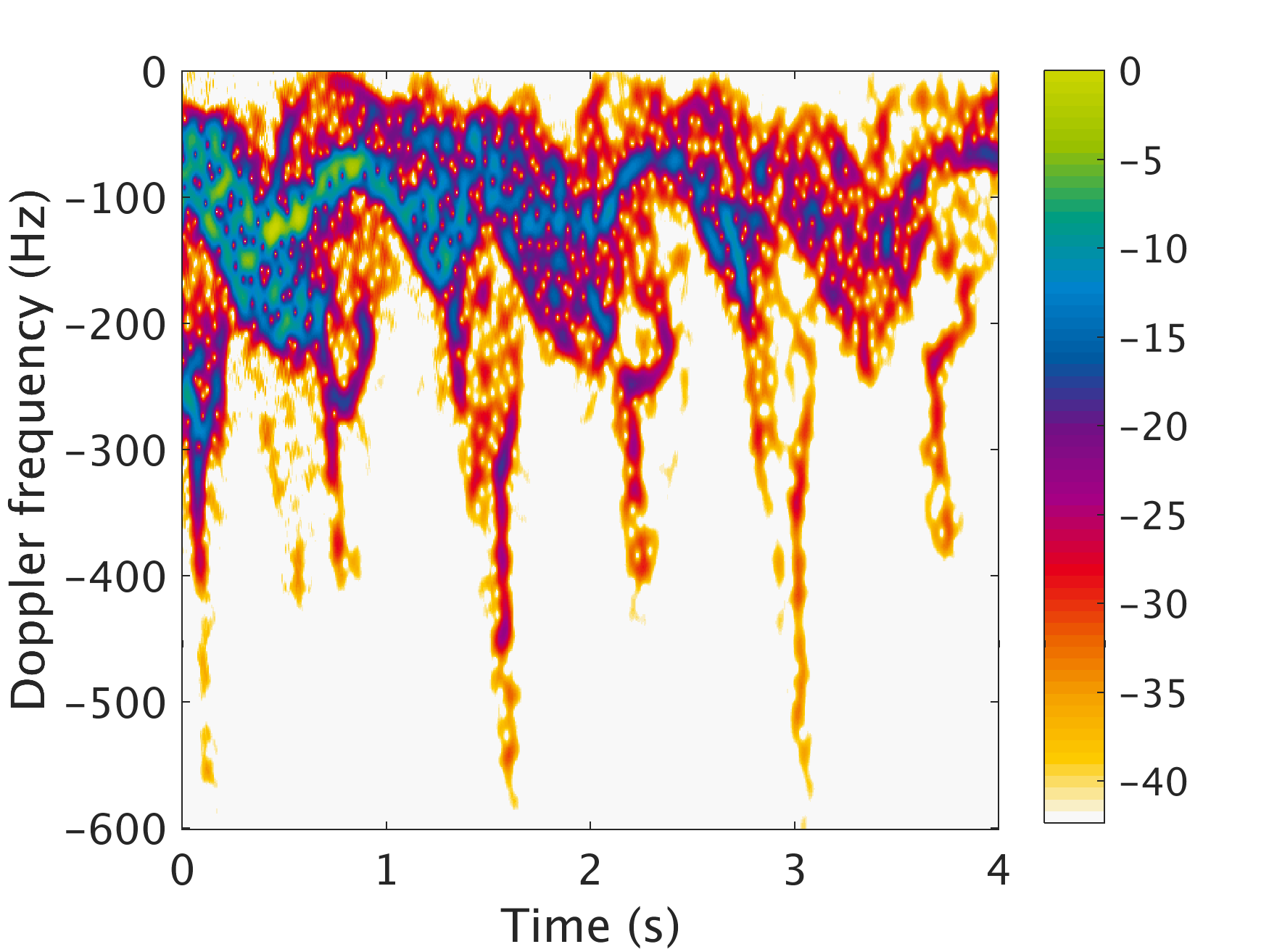}%
			\label{LB}}\vspace{-0.7em}}
	
	\centering{
		\subfloat[Person C]{\includegraphics[clip, trim= 0 0 20 18,width=0.5\columnwidth]{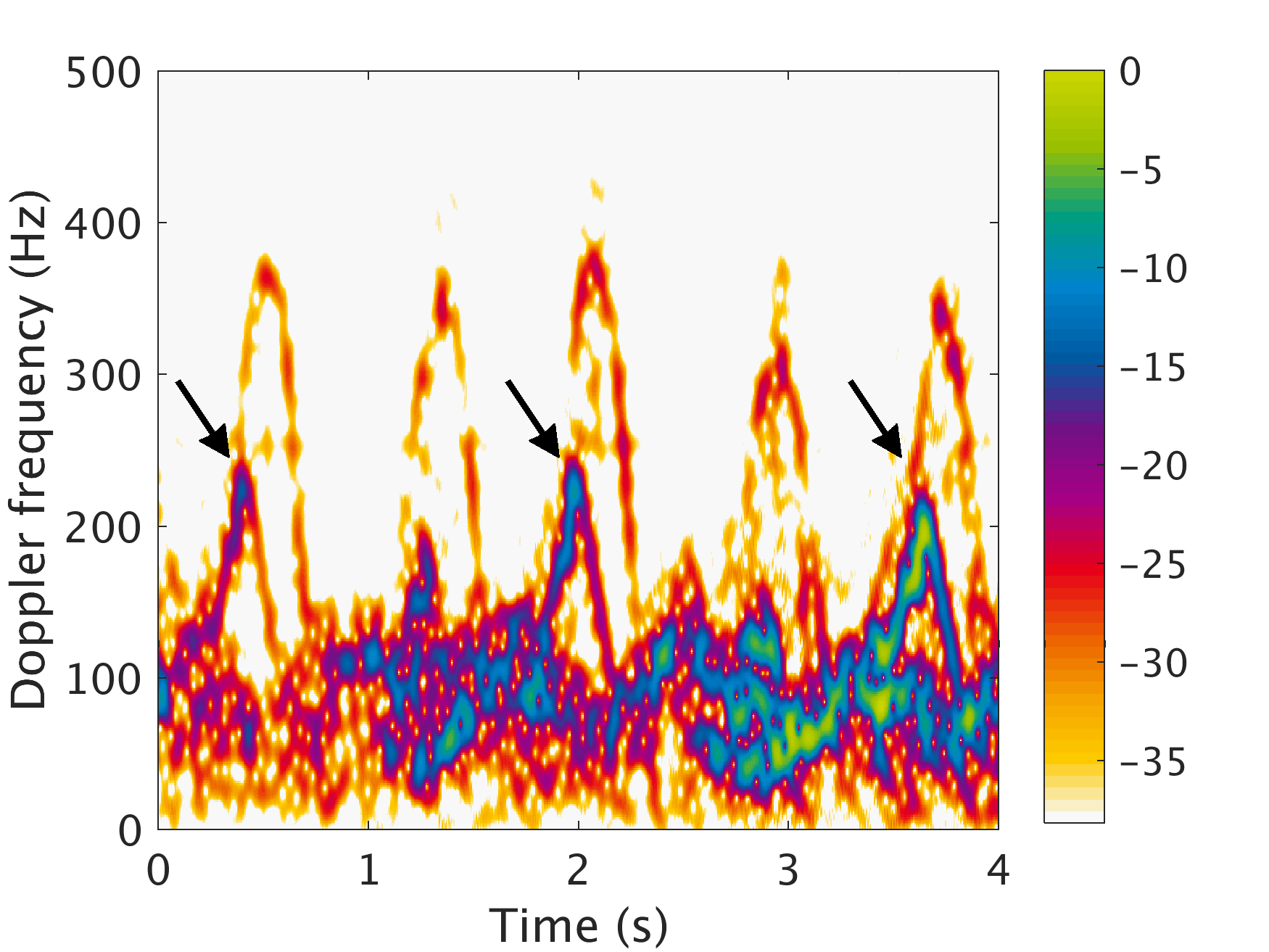}%
			\label{LC}}
		\subfloat[Person C walking with a cane]{\includegraphics[clip, trim= 0 0 20 18,width=0.5\columnwidth]{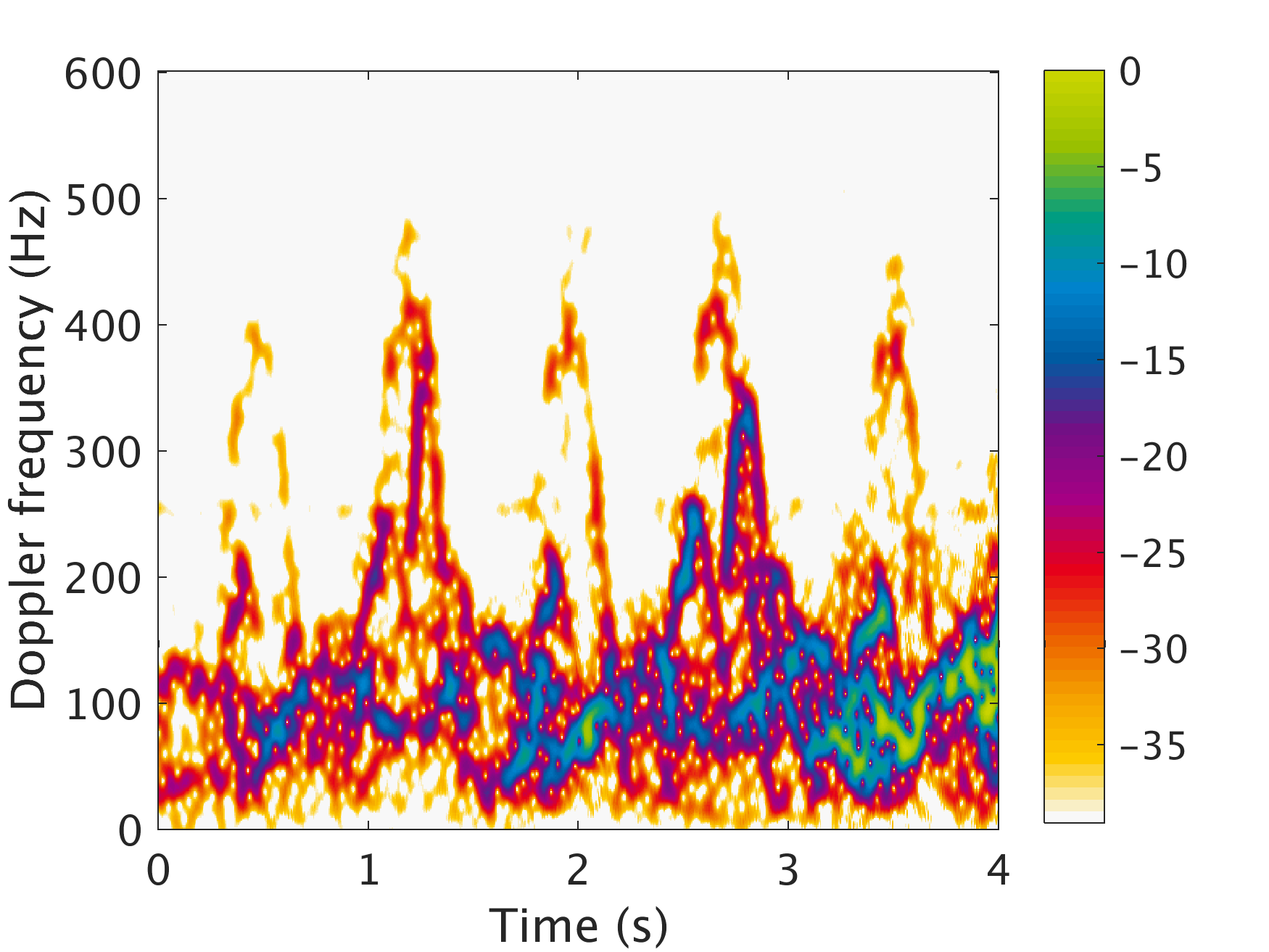}%
			\label{LCCane}}}
	\caption{Examples of spectrograms of four subjects with diagnosed gait disorders. The color indicates the energy level in dB.}
	\label{fig:specs_threesubjects}
	\vspace{-1em}
\end{figure}

In order to underscore the relevance of the acquired radar data, we also conducted experiments for radar data acquisition involving four test subjects with gait disorders due to different medical conditions. Examples of spectrograms for these subjects are shown in Fig.~\ref{fig:specs_threesubjects}. Figs.~\ref{fig:specs_threesubjects}\subref{LA}, \subref{LD} and \subref{LB} clearly show the same characteristic as the spectrogram of abnormal walking in Fig.~\ref{fig:specs}\subref{L1a}. Here, every other micro-Doppler stride signature has a lower maximal Doppler shift, which indicates an asymmetrical gait. In fact, as a result of a stroke at young age, Person A suffers from generalized dystonia affecting multiple muscle groups on one side of the body. Person B also experienced a stroke which caused a different gait disorder. In the case of Person C, due to the relative strength of the left side of the body, the asymmetry of the gait manifests itself in the knees' motions, rather than in different swinging velocities of the feet. Still, the spectrogram, as shown in Fig.~\ref{fig:specs_threesubjects}\subref{LC}, evidently reveals the gait asymmetry: on the onset of every other micro-Doppler stride signature, we can observe higher energy levels due to the altered stride motions (see arrows). Fig.~\ref{fig:specs_threesubjects}\subref{LCCane} shows a spectrogram of Person C walking with a cane, where the cane's signatures is overlapping with every other stride signature, similar to Fig.~\ref{fig:specs}\subref{CW}. The fourth person (D) has a congenital hip dislocation and suffers from a hip osteoarthritis on one body side due to it.

When applying the proposed classification method, i.e., using subspace-based features of pre-processed CVDs and the NN classifier, we can correctly identify the gait as abnormal in 92\% (12/13), 100\% (20/20), 75\% (9/12), and 100\% (26/26) of the cases for Person A, B, C, and D, respectively. The cane is correctly detected for Person C in 81\% (13/16) of the cases. Here, the classifier was trained based on the data of ten healthy individuals performing five different walking styles and evaluated using the data of the four individuals with pathological gait. Even though the observation time is only 6\,s per measurement, we can detect the asymmetry of the gait with very high sensitivity (TPR). These results, which are based on Doppler radar data representations of subjects with diagnosed gait disorders, are very promising and will serve as a basis for more extensive studies. 
\section{Conclusion}\label{sec:conclusion}
In this paper, different walking styles were analyzed based on radar micro-Doppler signatures and their Fourier transforms. Methods were presented to perform gait recognition utilizing the cadence-velocity domain. These methods include data-driven feature learning and features related to motion kinamatics. Subspace features were superior to standard physical features in solving the intra motion category classification problem of discerning different gaits. Experimental results have shown that five simulated gait classes can be identified on a small population of healthy subjects and patients. In particular, the gait abnormality of four individuals with diagnosed gait disorders was correctly identified with high sensitivity. Future work should consider a wider group of patients with pathological gait.




\bibliographystyle{IEEEtran}
\bibliography{ref_akseifert}

\end{document}